\documentclass[10pt,preprint]{sensys-proc}

\usepackage{subfloat}
\usepackage{color}
\usepackage{ifthen}
\usepackage{soul}
\usepackage{graphicx}
\usepackage{balance}
\usepackage{comment}
\usepackage{amssymb}
\usepackage{amsmath}
\usepackage{gensymb}
\usepackage[hyphens]{url}
\usepackage{flushend}

\newboolean{showComments}
\setboolean{showComments}{true} 
\ifthenelse{\boolean{showComments}}{%
\newcommand{\cm}[1]{\sethlcolor{yellow}\hl{[Cecilia: #1]}}
\newcommand{\sn}[1]{\sethlcolor{green}\hl{[\emph{Sarfraz:} #1]}}
\newcommand{\todo}[1]{\sethlcolor{cyan} \hl{[ToDo: #1]}}
}{%
\newcommand{\cm}[1]{}
\newcommand{\sn}[1]{}
\newcommand{\todo}[1]{} 
}

\newcommand{\hidden}[1]{}

\numberofauthors{2}

\author{
\alignauthor Sarfraz Nawaz \\
     \affaddr{Computer Laboratory}\\
     \affaddr{University of Cambridge}\\ 
      \email{sarfraz.nawaz@cl.cam.ac.uk} \\            
\alignauthor Cecilia Mascolo \\ 
    \affaddr{Computer Laboratory}\\
    \affaddr{University of Cambridge}\\
    \email{cecilia.mascolo@cl.cam.ac.uk}        
}

\crdata{978-1-4503-3143-2}
\conferenceinfo{SenSys'14,} {November 3--6, 2014, Memphis, TN, USA.}
\CopyrightYear{2014}

\title{Mining Users' Significant Driving Routes \\
with Low-power Sensors}

\begin{document}

\maketitle

\begin{abstract}

While there is significant work on sensing and recognition of significant places for users, little attention has been given to users' {\em significant routes}. Recognizing these routine journeys, opens doors to the development of novel  applications, like personalized travel alerts, and enhancement of user's travel experience. However, the high energy consumption of traditional location sensing technologies, such as GPS or WiFi based localization, is a barrier to passive and ubiquitous route sensing through smartphones.

In this paper, we present a passive route sensing framework that continuously monitors a vehicle user {\em solely through a phone's gyroscope and accelerometer}. This approach can differentiate and recognize various routes taken by the user by time warping angular speeds experienced by the phone while in transit and is independent of phone orientation and location within the vehicle, small detours and traffic conditions. We compare the route learning and recognition capabilities of this approach with GPS trajectory analysis and show that it achieves similar performance. Moreover, with an embedded co-processor, common to most new generation phones, it achieves energy savings of an order of magnitude over the GPS sensor.




\end{abstract}




\section{Introduction}

Modern smartphones have a wide range of embedded sensors and are increasingly being used as a novel sensing platform to sense all aspects of a user's life ranging from health and well-being to driving style. They are a perfect tool for sensing user's context and proactively providing information to the user that he or she will find useful like a virtual personal assistant~\cite{googlenow}. One form of this context sensing is \emph{place learning} that identifies significant places where a user usually spends some of his or her time regularly, for example, home, work or lets say a gym. While there is significant amount of work on learning these significant places~\cite{senseloc, learningplaces, csp}, little attention has been given to {\em significant journeys}, i.e., journeys that users regularly make during their daily routines. 

The detection of a user's significant journeys can have various applications in terms of the general quantified self movement: mapping the daily displacements of a person as well as his or her activity has become a fashionable demand in recent years. In addition, the early detection of the fact that a user has started routine journey can improve  travel recommendations and traffic updates. We have also seen applications related to preparation of the destination location for user's arrival or departure: smart home heating systems~\cite{tado} switching on/off when a journey to/from a place is detected, social life alerts to family of the arrival of person, and so on.

However, continuous monitoring of user's journeys can be expensive as location sensing through user's personal device can incur prohibitive energy costs. While it is possible to use duty cycling~\cite{jigsaw} or trigger location sensing~\cite{rateadaptive} from low power sensors, the energy cost of location sensing is high when location sensors are active. It is also possible for the phone to be plugged in a charging port in the vehicle during journeys but this places an extra constraint on the user to remember to plug the phone in during each journey for energy intensive location sensing. Significantly lowering the energy consumption of sensing these journeys will relieve the user of this requirement and thus make it more likely and convenient for users to adopt and use the system. Therefore, in this paper, we explore the use of alternative low energy phone sensors for accurate detection of user's significant routes. Sensors such as accelerometer and gyroscopes can be reasonably cheap for the detection of user patterns such as change in user activity and means of travel. A large body of literature has covered user activity detection~\cite{activity1, activity2, activity3}, transport mode detection~\cite{acceltransport, transportmodes} and other vehicle and travel related applications~\cite{pothole, pothole2, wreckwatch, drunkdriving, vehicledynamics} using these sensors. The use of these sensors has become even more energy efficient recently, thanks to the advent of embedded co-processors in modern phones such as the iPhoneÕs M7 motion co-processor and QualcommÕs Hexagon QDSP6 of the Snapdragon 800 processor platform available on Google Nexus 5, Nokia Lumia 1520, Sony Xperia Z1 and Samsung Galaxy Note 3 (LTE).

In this paper, we take advantage solely of accelerometer and gyroscope of a user's phone to detect with high accuracy significant, i.e. repeated, user journeys. The approach is able to distinguish user routes by employing Dynamic Time Warping of angular speeds experienced by the phone while in transit and is independent of phone's  orientation, location within the vehicle and traffic conditions. In order to show the effectiveness of the approach we compare this route learning framework with one where GPS is used and showing comparable performance if only the CPU is used for sensing and processing and then with an approach which takes advantage of co-processor computation achieving energy savings of an one order of magnitude over the GPS sensor.

To our knowledge, this is the first approach of its kind. While inertial navigation can be used to track the position of a vehicle using inertial sensors like accelerometer, gyroscope and magnetometer, the noise characteristics of low cost MEMS inertial sensors~\cite{inertialtr, bumping} in smartphones make them unsuitable for inertial only navigation and tracking. A MEMS sensor based inertial navigation system either requires periodic external information~\cite{bumping, autowitness} or extensive user training and calibration~\cite{acc} to be useful.

\begin{figure}[t!]
\centering
\includegraphics[width=0.9\columnwidth]{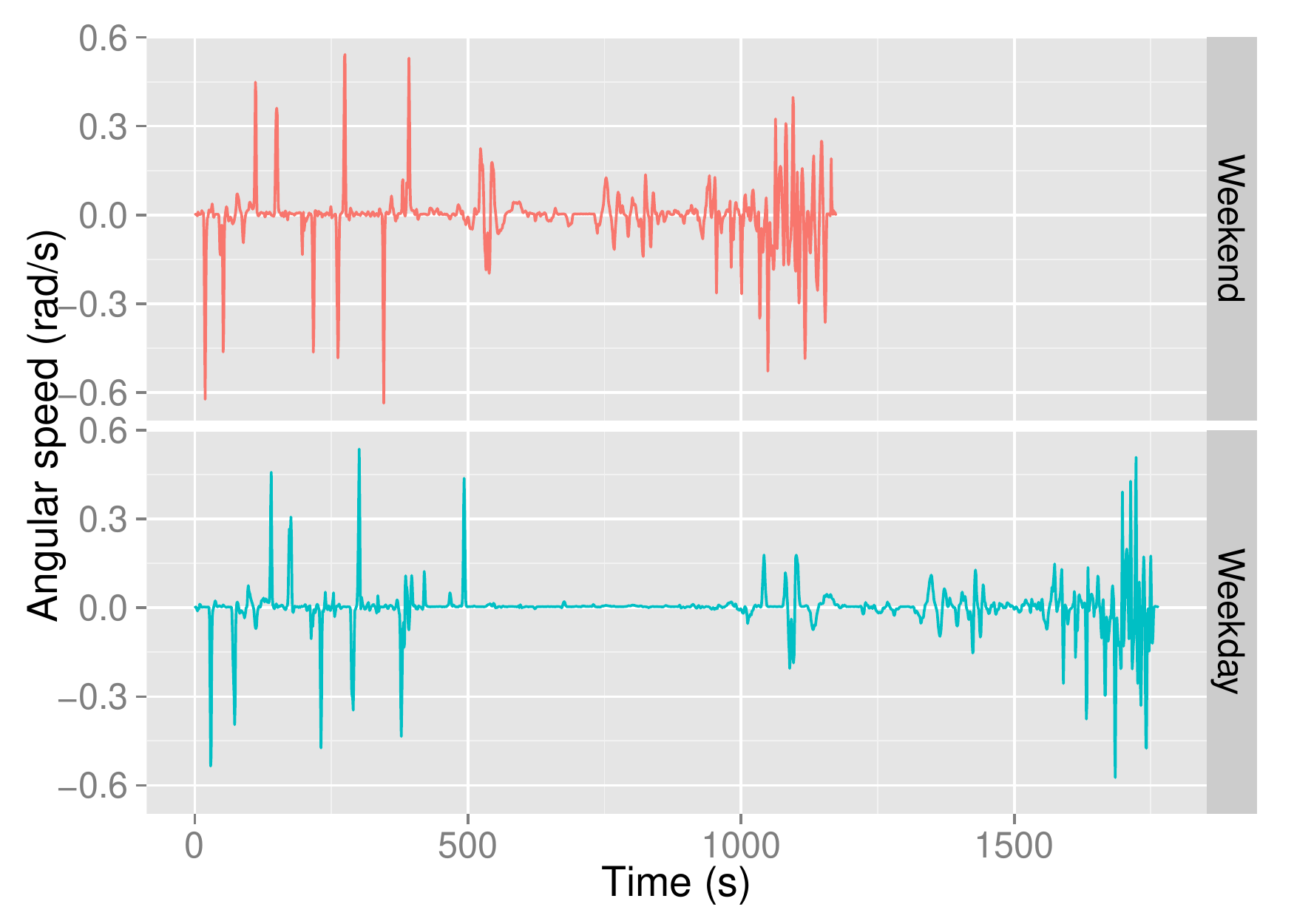}
\caption{Comparison of processed gyroscope output from a smartphone on two different days for the same route.}
\label{fig:gyroexample}
\end{figure}

To summarize, the contributions of this work are the followings
\begin{itemize}
\item A Dynamic Time Warping based framework for significant route detection based solely on accelerometer and gyroscope sensors in the phone.
\item A comparison of the sensing performance against a GPS based approach. The results show comparable performance.
\item A prototype implementation of the system with a low power co-processor.
\item A comparison of the energy consumption of co-processor based system with the GPS based approach. The results show one order of magnitude of energy savings over the GPS based approach.
\end{itemize}
%
%
%

\section{Approach Overview}
\label{sec:approachoverview}
\begin{figure}[t!]
\centering
\begin{minipage}[b]{0.47\linewidth}
  \includegraphics[scale=0.15]{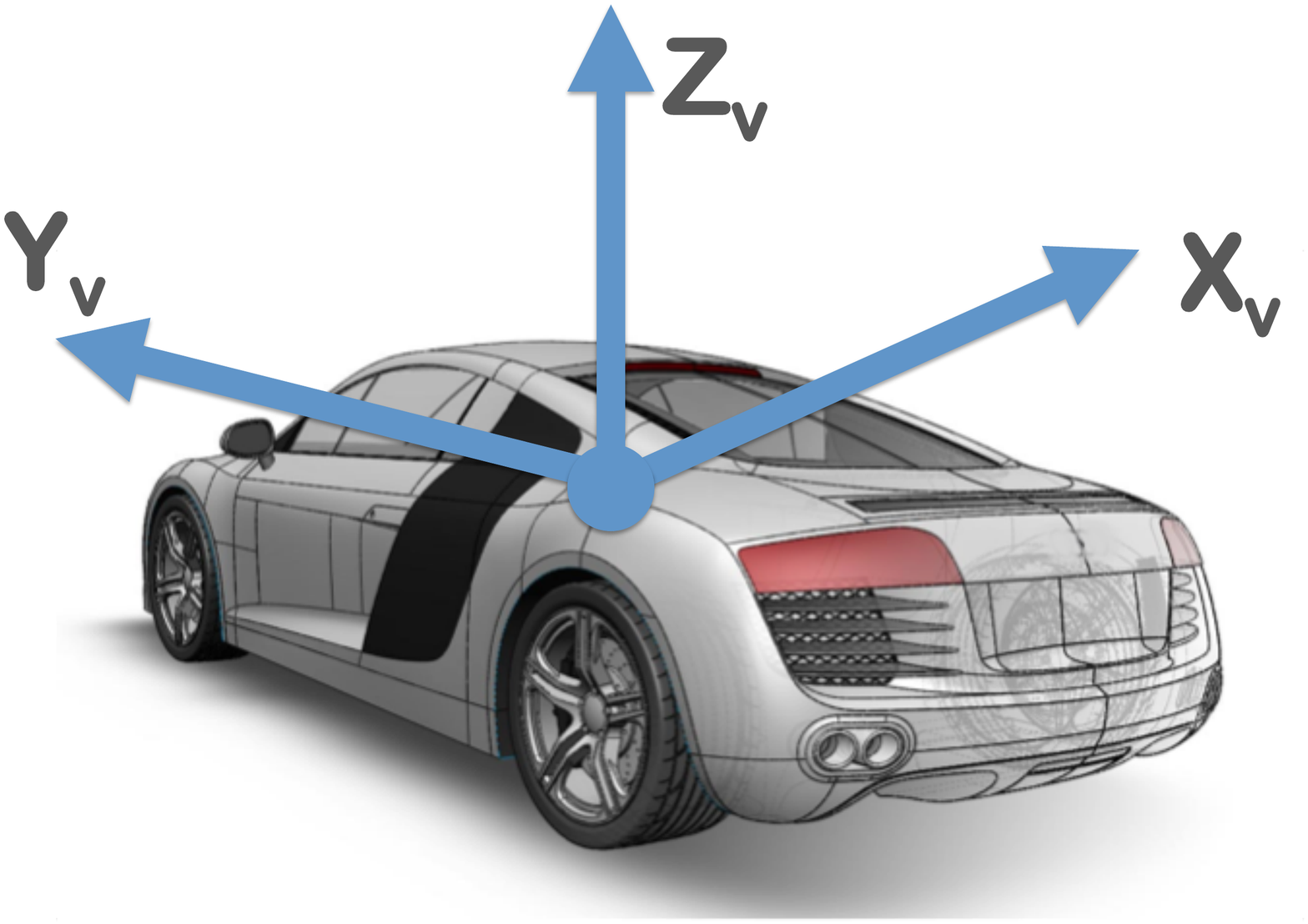}
  \caption{Vehicle coordinate system}
  \label{fig:vcoords}
\end{minipage}
\begin{minipage}[b]{0.47\linewidth}
  \includegraphics[scale=0.15]{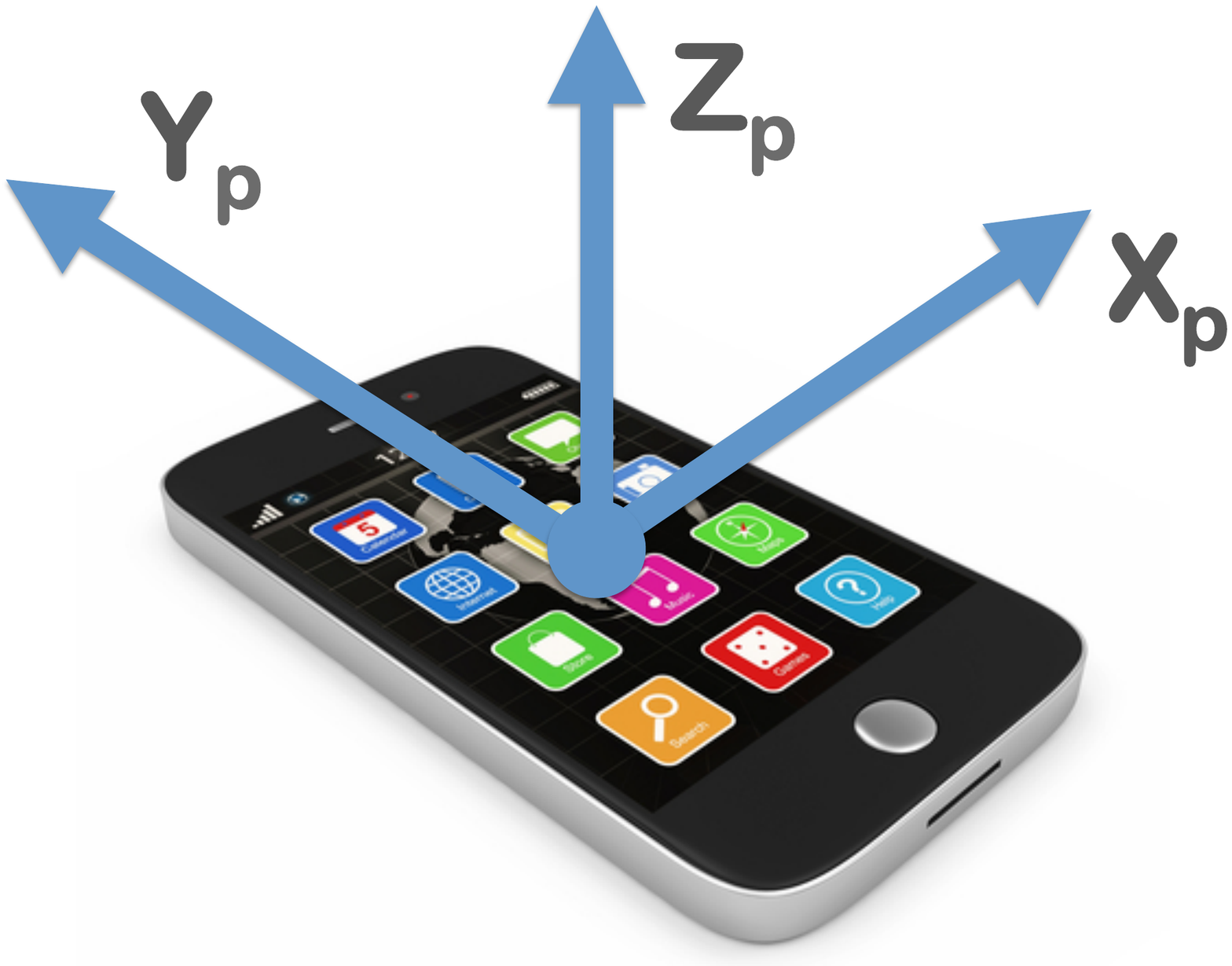}
  \caption{Phone coordinate system}
  \label{fig:pcoords}
\end{minipage}
\end{figure}

In this section we describe how our approach of sensing significant driving routes takes advantage of a phone's sensors. When a user carrying a smartphone travels in a car, the sensors embedded in the phone, for example, the accelerometer and the gyroscope experience different forces. The accelerometer can sense the acceleration and deceleration of the vehicle and the gyroscope can sense that the vehicle is turning~\cite{vehicledynamics}. As the vehicle travels, each turn and bend of the road causes a change in the angular speed sensed by the gyroscope. This creates a unique \emph{signature} of the journey in the form of variations of angular speed. When the user repeats this journey, the phone sensors can recognize the route by observing that the variations in angular speed are similar to the ones of the previous journey. For example, the sequence of turns that a vehicle takes to get from one location to a destination remains the same between different journeys on the same route. 

\begin{figure*}[t]
\centering
\begin{minipage}[b]{0.3\linewidth}
  \includegraphics[scale=0.3]{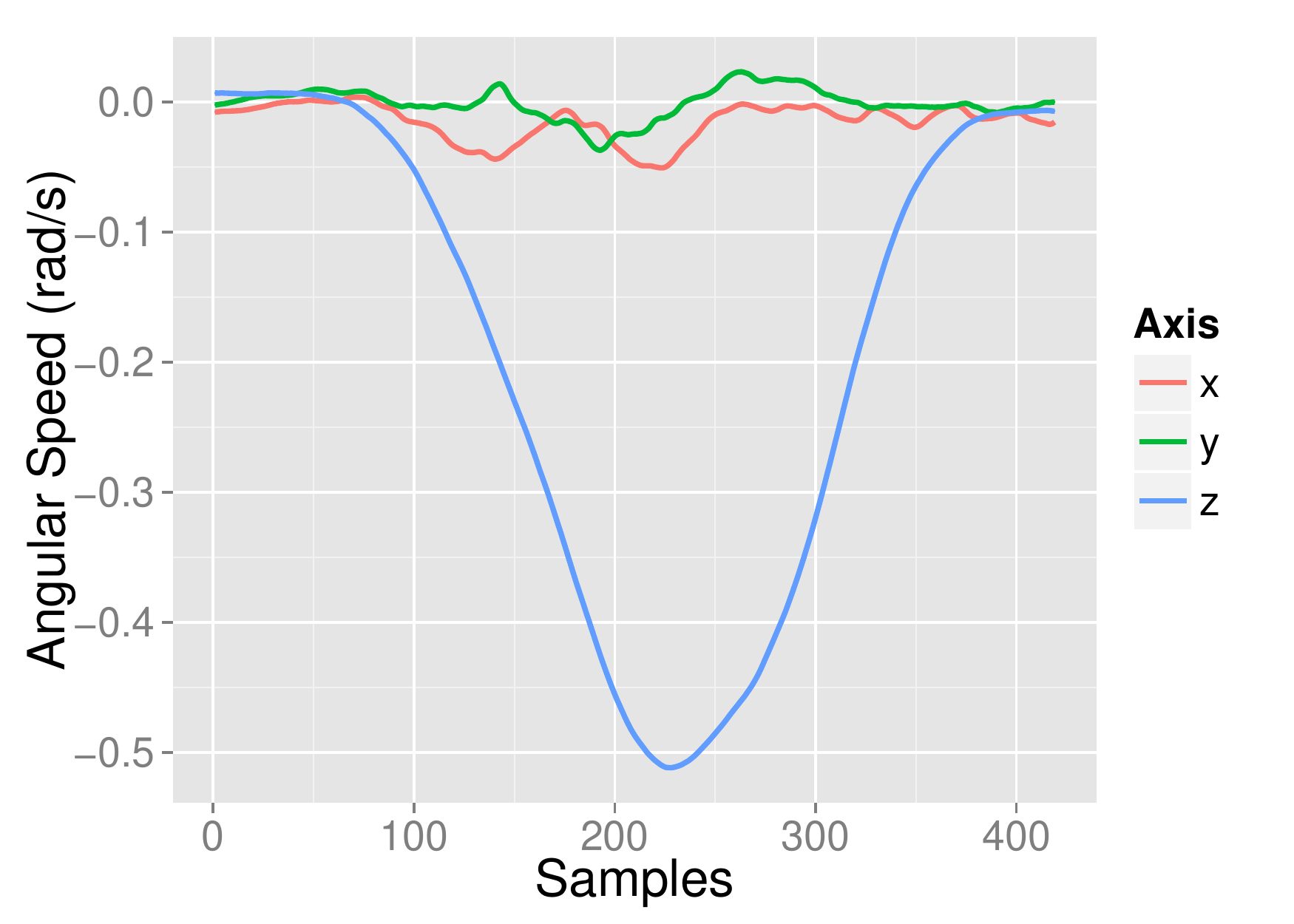}
  \caption{Vehicle aligned}
  \label{fig:turnaligned}
\end{minipage}
\begin{minipage}[b]{0.3\linewidth}
  \includegraphics[scale=0.3]{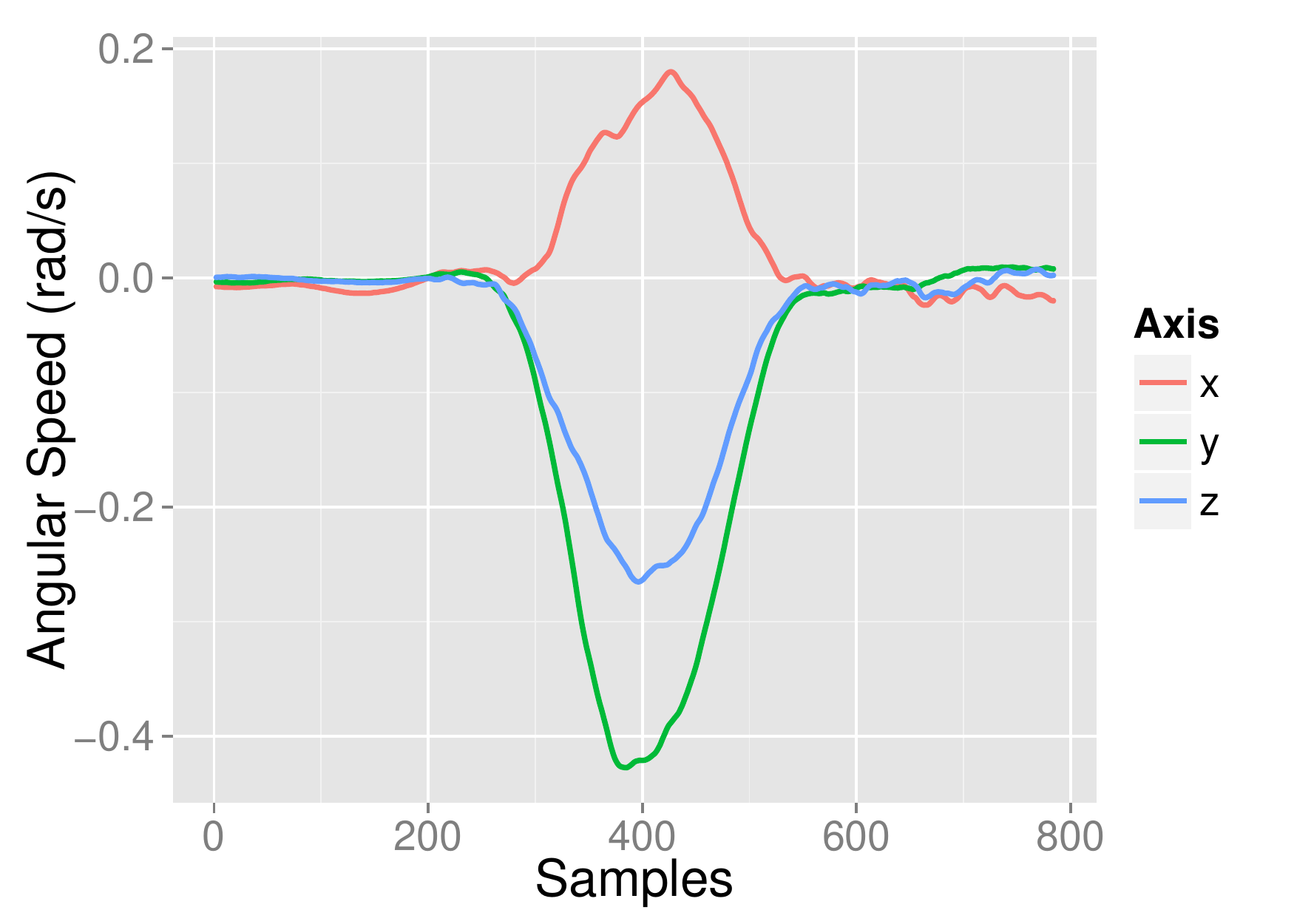}
  \caption{Arbitrary orientation}
  \label{fig:turnunaligned}
\end{minipage}
\begin{minipage}[b]{0.3\linewidth}
  \includegraphics[scale=0.3]{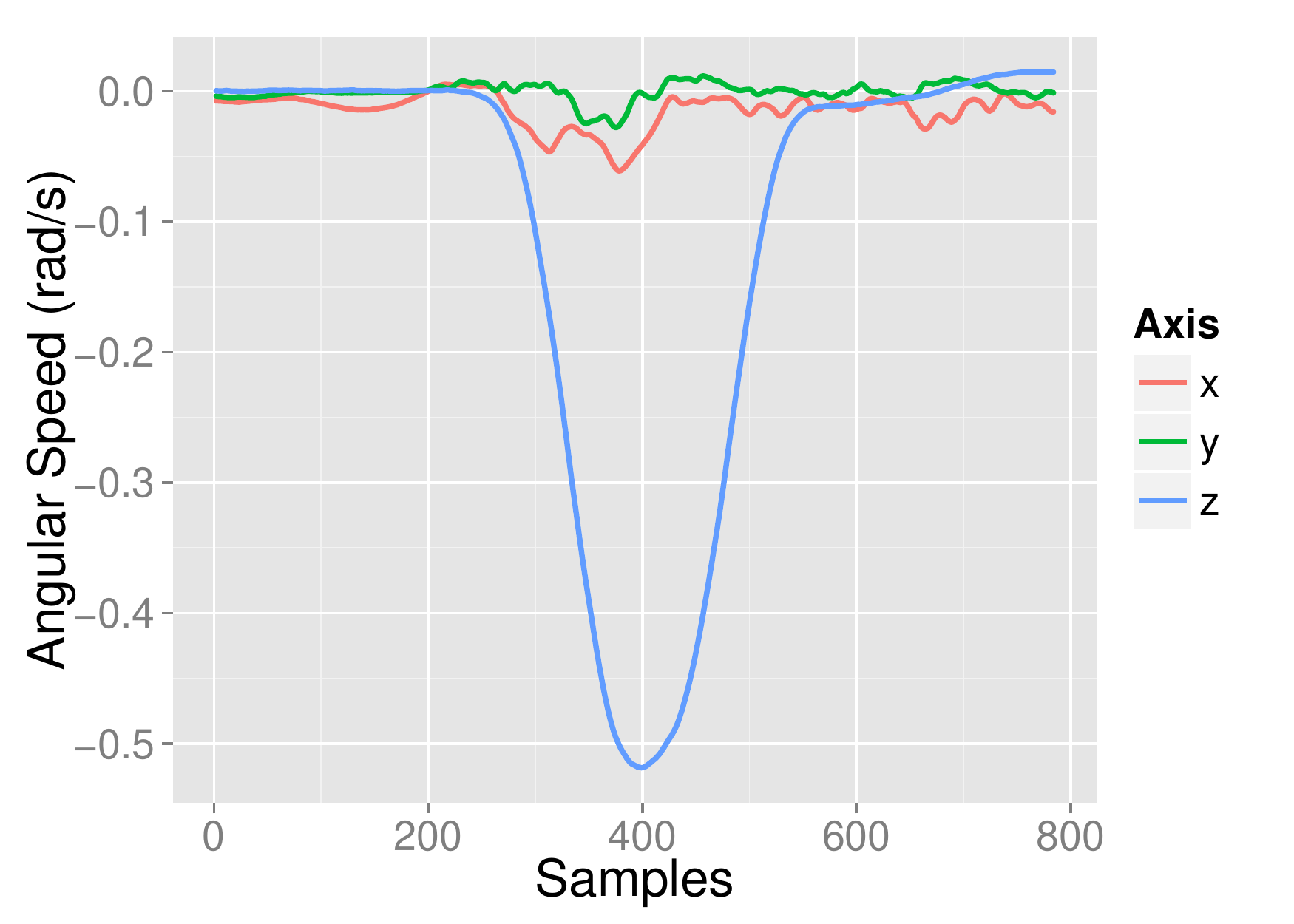}
  \caption{After Z-Alignment}
  \label{fig:turnzaligned}
\end{minipage}
\end{figure*}

This observation forms the basis of this work. But how repeatable are these patterns and can we really recognize repeated routes from these patterns? Fig. (\ref{fig:gyroexample}) shows two such angular speed patterns captured by the gyroscope of a phone carried in a  vehicle from a $12$km long route (shown in Fig. \ref{fig:paths} as actual route) in a typical urban city when the route is driven on on two different days. The output of the gyroscope has been low-pass filtered and processed to account for phone's orientation in the vehicle. We will discuss the details in Section \ref{sec:zaxis_alignment}. The upper red trace is from a journey that took place on a weekend and the lower blue trace is from a second journey on the same route in the weekday evening traffic. Two things are evident from this figure: 1) the patterns bear a strong similarity; and 2) these are stretched (or compressed) in time with non-uniform scaling on the time axis. There are some subtle differences too, i.e., some of the angular speed peaks have slightly different magnitudes. The angular speed patterns are similar because the vehicle takes the same route and thus executes same sequence of turns and bends. But since the vehicle travels at different speeds on different parts of the route depending on traffic, this leads to non-uniform scaling of the pattern on the time axis. And as the vehicle can take same turns at different speeds on different occasions, the magnitudes of some of the peaks can be slightly different. Our approach is able to  compare the overall similarity of these two patterns under non-uniform time scaling while accounting for the arbitrary way a smartphone can be carried in the vehicle so to recognize repeated routes from the gyroscope sensor in the phone. In the next section we discuss in detail how this can be achieved.

\section{Approach in Detail}
\label{approach}

In this section we describe the details of our approach. We first show how we take into account that a user's phone will be oriented arbitrarily. We then describe the dynamic time warping approach which is at the base of the alignment of the journey traces collected with the sensors.

\subsection{Z-Axis Alignment}
\label{sec:zaxis_alignment}

A user usually carries a phone in an arbitrary orientation while traveling in a vehicle. This means that our system cannot use the inertial sensors in the phone to sense turns unless we align the coordinate system used by the phone to a certain orientation. Here we describe how we address this issue. Let us first define two coordinate systems, a vehicle coordinate system and a phone coordinate system. The vehicle coordinate system is defined by three orthogonal axes $X_v$, $Y_v$ and $Z_v$ where $+Y_v$ points in the forward traveling direction, $+Z_v$ points towards the sky and $+X_v$  extends to the right as shown in the Fig. (\ref{fig:vcoords}). Similarly the phone coordinate system is defined by orthogonal axes $X_p$, $Y_p$ and $Z_p$ with positive directions as shown in Fig. (\ref{fig:pcoords}). As a vehicle takes a turn, it rotates on its $Z_v$ axis but there is almost no rotation around $X_v$ or $Y_v$. This rotation can be sensed by the gyroscope of a phone in the vehicle. Fig. (\ref{fig:turnaligned}) shows the output from the gyroscope as the vehicle carrying the phone takes a right turn. The coordinate system of the phone is aligned to that of the vehicle. It shows that during the turn almost all of the angular speed is contained in the $z$ component whereas the $x$ and $y$ components are negligible because there is little rotation around these axes. This means that we only have to align $Z_p$ axis of the phone with $Z_v$ axis of the vehicle to capture its rate of rotation during turns. 


Now we describe how we can align $Z_p$ with $Z_v$. $Z_v$ is in fact aligned with gravity and we can easily estimate the inclination angle $\theta$ between $Z_p$ and $Z_v$ using the accelerometer.  Let $\mathbf{g_p} = [a_x, a_y, a_z]^T$ be a vector estimate of gravity derived from the accelerometer.  It is obtained by running a low pass filter on accelerometer output to extract the DC component in the signal~\cite{androidgravity, lowpass}. If $\mathbf{\hat{g}_p}$ is a unit vector along $\mathbf{g_p}$ then we can calculate a quaternion $\mathbf{q}$ ~\cite{quatintro} that rotates $Z_p$ to $Z_v$ through the shortest arc as,

\begin{equation}
	\mathbf{q} = \cos\frac{1}{2}\theta + \mathbf{\hat{u}}\sin\frac{1}{2}\theta 
\end{equation}
\begin{eqnarray}
	\mathbf{u} &=& \mathbf{\hat{g}_p}\times\mathbf{\hat{g}} \\
	\mathbf{\hat{u}} &=& \frac{1}{\lVert\mathbf{u}\rVert}\mathbf{u} \\
	\nu &=& \mathbf{\hat{g}_p}.\mathbf{\hat{g}} \\
	\cos\frac{1}{2}\theta &=& \sqrt{\frac{1+\nu}{2}} \\
	\sin\frac{1}{2}\theta &=& \sqrt{\frac{1-\nu}{2}} 
\end{eqnarray}
where $\mathbf{u}$ and $\nu$ are cross and dot products respectively and $\mathbf{\hat{g}} = [0, 0, 1]^T$. Now we can rotate  $Z_p$ to align it with $Z_v$ using Hamilton product~\cite{quatintro} and calculate the aligned output of the gyroscope $\mathbf{\omega}_r$ as,

\begin{equation}
	\mathbf{\omega}_r=\mathbf{q}\mathbf{\omega}\mathbf{q}^{-1}
	\label{eq:rotatedgyro}
\end{equation}
where $\mathbf{\omega} = [\omega_x, \omega_y, \omega_z]^T$ is the output of the gyroscope and $\mathbf{q}^{-1}$ is conjugate quaternion,
\begin{equation}
	\mathbf{q}^{-1} = \cos\frac{1}{2}\theta - \mathbf{\hat{u}}\sin\frac{1}{2}\theta 
\end{equation}

\begin{figure}[t]
\centering
\includegraphics[scale=0.33]{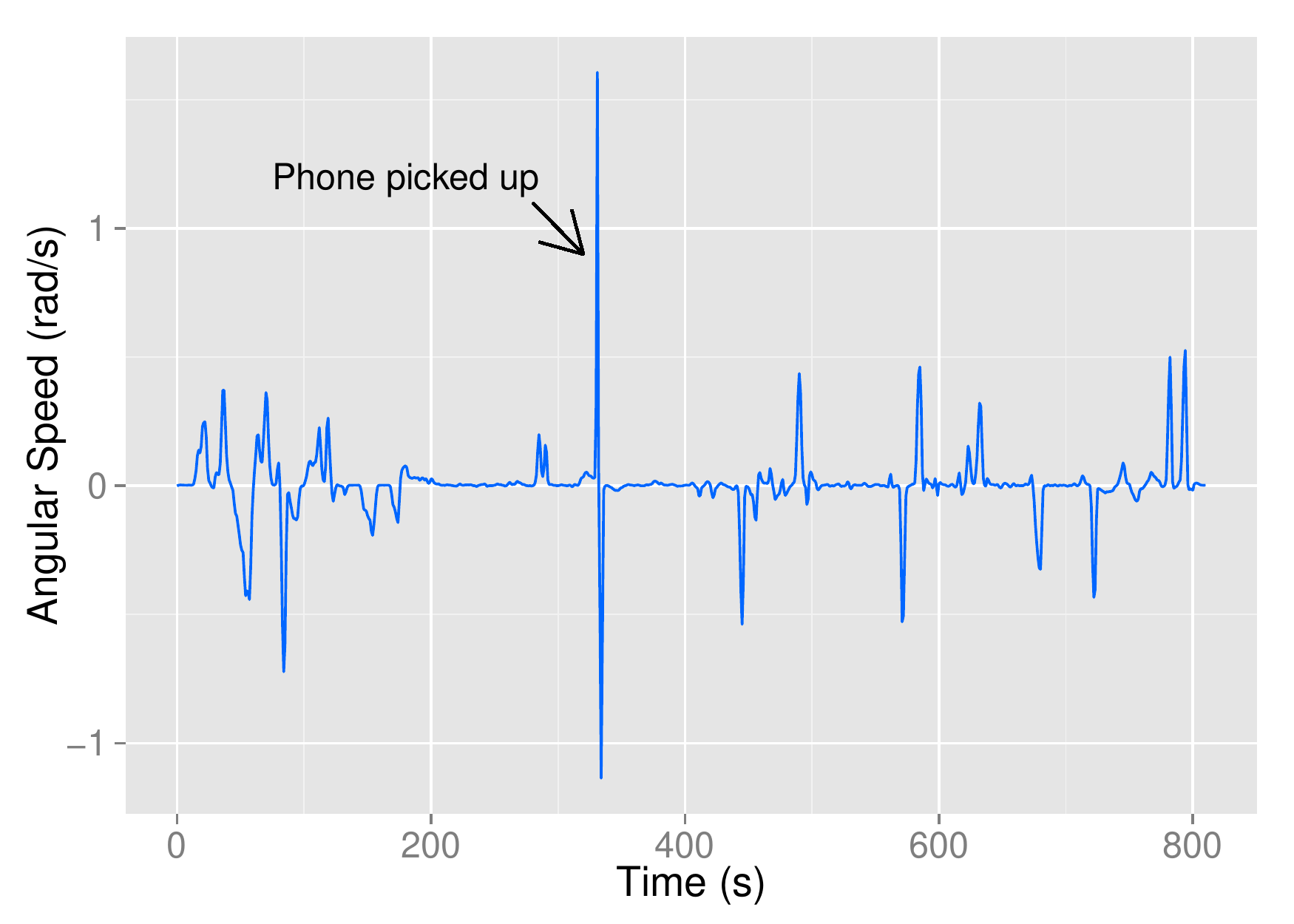}
\caption{Detecting user interaction with the phone}
\label{fig:picked}
\end{figure}

Here we must point out that we do not align the two coordinate systems completely as in other sensing systems~\cite{nericell, vehicledynamics}. We only align the $z$ axis of the phone with the $z$ axis of the vehicle. The $x$ and $y$ axes of the phone, $X_p$ and $Y_p$ could be pointing in any direction. However, this does not pose any problems for our application because we only want to sense the angular speed of the vehicle which is contained almost entirely in the z component. In order to demonstrate this we collect gyroscope data from a phone with an arbitrary orientation in the vehicle as it turns right. Fig. (\ref{fig:turnunaligned}) shows the output of the gyroscope. In this case, none of the phone axes are aligned with any of the vehicle axes and this results in components of total angular speed appearing on all three axes of the phone. Fig. (\ref{fig:turnzaligned}) shows the gyroscope output after it has been rotated using Eq. (\ref{eq:rotatedgyro}) of our Z-axis alignment approach. Comparing Fig. (\ref{fig:turnzaligned}) with Fig. (\ref{fig:turnaligned}) shows that it is sufficient to align $Z_p$ with $Z_v$ to sense vehicle's angular speed and therefore the turns it makes during a journey. Our application and the proposed approach, therefore, does not require complete coordinate alignment. While traveling in a vehicle, users typically carry the phone in their clothing, in a bag or carefully place it in a secure location in the vehicle to avoid any damage. It is, therefore, unlikely for the phone to experience continuous random perturbations during the journey. However, if the user interacts with the phone during the journey, it changes device orientation. But it is easy to detect such an interaction because it causes a large change in gyroscope output. Fig. (\ref{fig:picked}) shows the gyroscope output from a journey where the user picked up the phone while driving and then placed it back. We can use a threshold to detect such an interaction and perform z-axis alignment again. 

\subsection{Dynamic Time Warping}
\label{sec:dtw}

Dynamic time warping is a class of algorithms that calculate dissimilarity or distance between two ordered sequences while allowing compression or expansion of the ordered axis to best align the two sequences. These algorithms map each point in one sequence to one or more points in the second sequence using a dynamic programming approach. Dynamic time warping (DTW) was originally proposed for speech recognition~\cite{dtwspeech1, dtwspeech2} to account for different speaking rates by different speakers. However, since then it has been extensively used for time series analysis, clustering and classification in bioinformatics~\cite{biodtw}, shape recognition~\cite{shapedtw}, gesture recognition~\cite{gesturedtw} and many other applications.

\begin{figure}[t]
\centering
\includegraphics[scale=0.3]{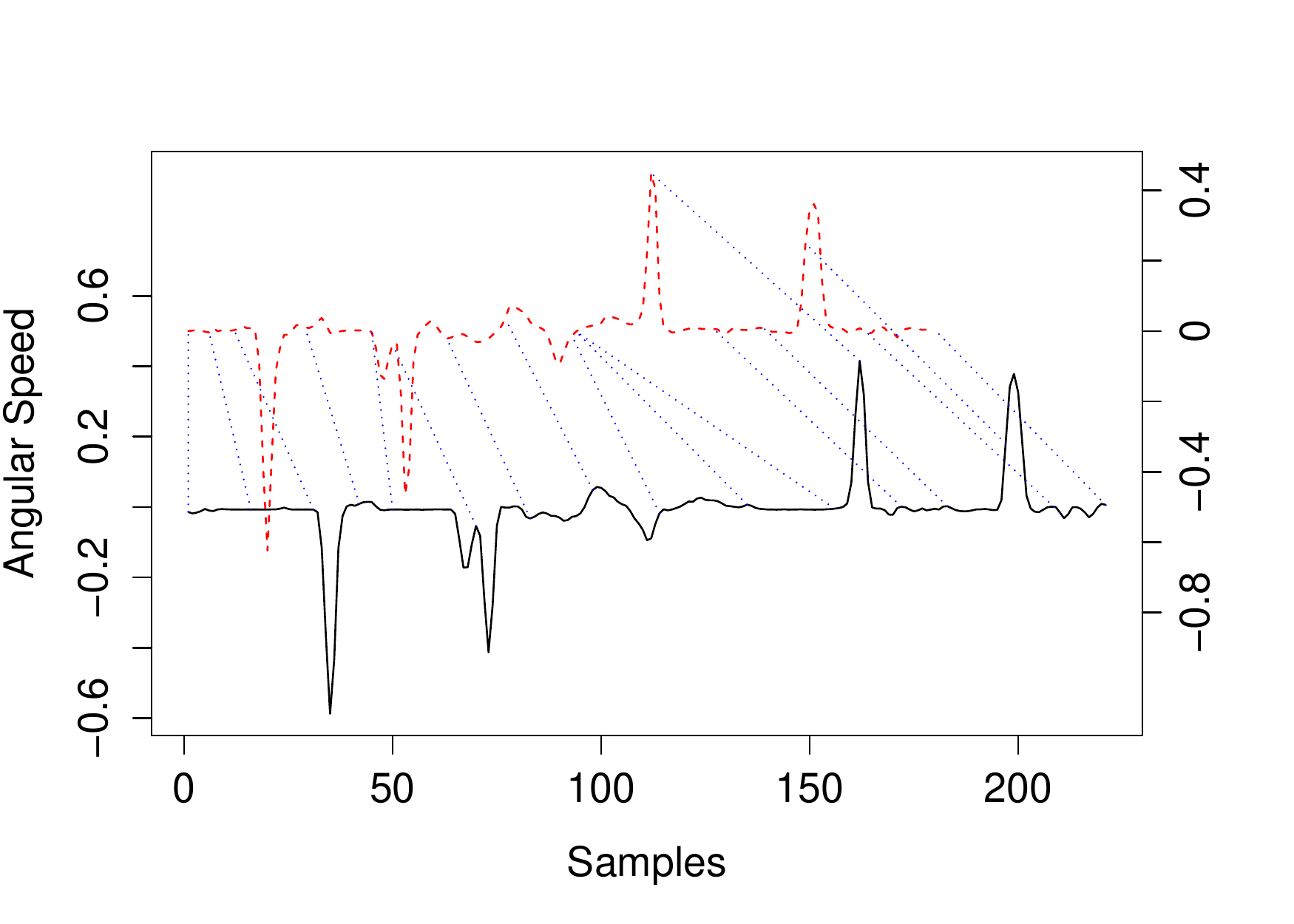}
\caption{Dynamic time warping}
\label{fig:dtw}
\end{figure}

Let us suppose that we have a time series of angular velocities $X = \lbrace x_1, x_2, x_3 \hdots, x_n \rbrace$ from the gyroscope that has already been aligned with the z-axis of the vehicle. And we want to compare it to a second time series $Y = \lbrace y_1, y_2, y_3 \hdots, y_m \rbrace$ that was captured from a previous journey. If $i = 1, 2, 3, \hdots, n$ and $j = 1, 2, 3, \hdots m$ are indices into $X$ and $Y$ respectively and $P$ is an optimal mapping  between the two sequences given as,

\begin{eqnarray}
	P(k) &=& (p_x(k), p_y(k)) \\
	k &=& 1, 2, 3, \hdots T \nonumber \\
	p_x(k) &\in& \lbrace 1, 2, 3, \hdots, n\rbrace \nonumber \\
	p_y(k) &\in& \lbrace 1, 2, 3, \hdots, m\rbrace \nonumber
\end{eqnarray}
we can compute the dissimilarity or distance between the two sequences as,

\begin{equation}
	d_{DTW}(X,Y) = \sum_{k=1}^Td(p_x(k), p_y(k))m(k)/M
	\label{eq:dtwnormalized}
\end{equation}
where $m(k)$ is a per step weight associated with the step pattern discussed later, $M$ is a normalization constant to account for different lengths of $X$ and $Y$ and $d(i, j)=f(x_i, y_i) \geq 0$ is a non-negative function to compute dissimilarity between individual elements of $X$ and $Y$. The mapping $P$ is also called a warping path and is calculated by solving the following problem under certain constraints.

\begin{figure}[t]
\centering
\includegraphics[scale=0.36]{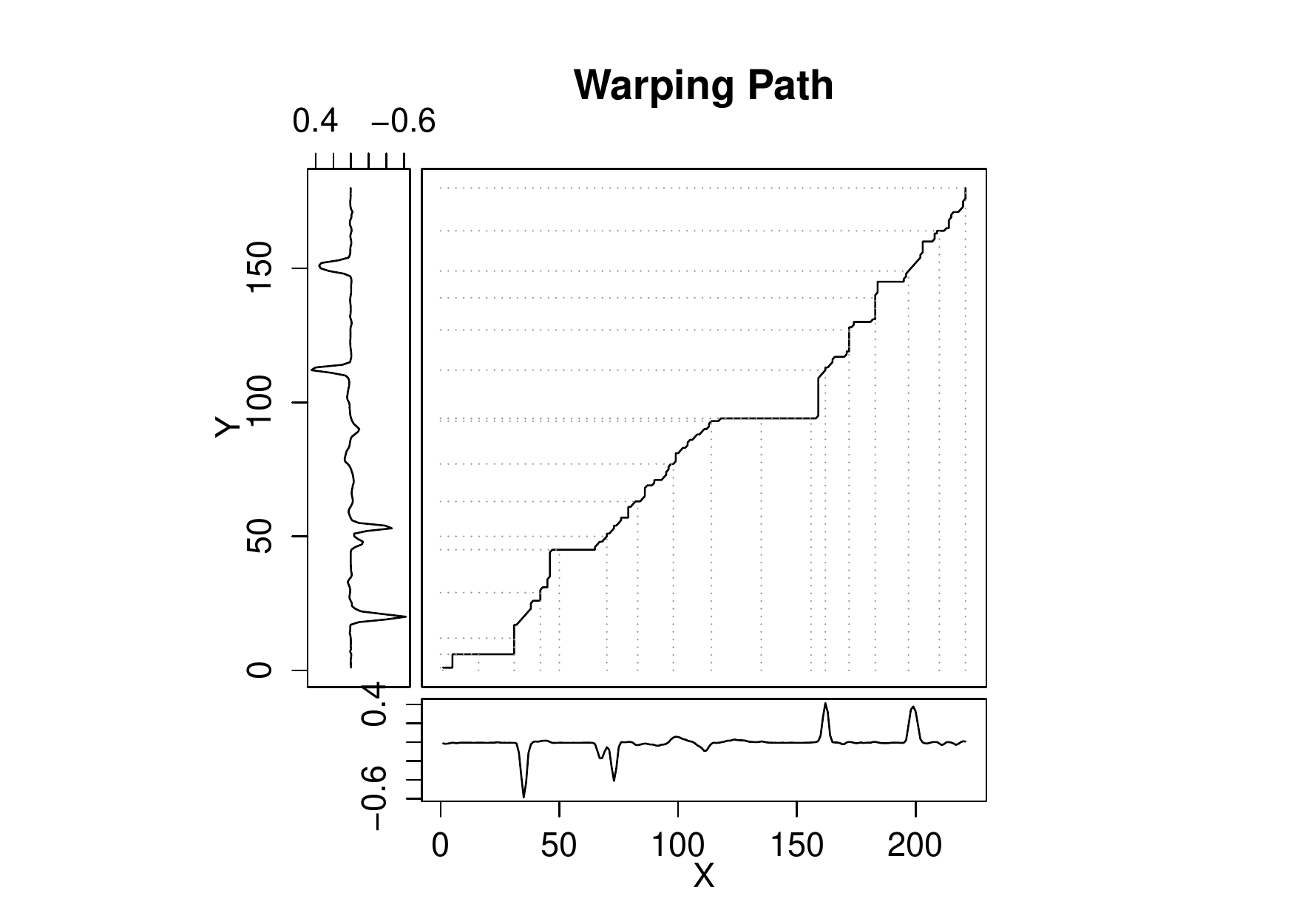}
\caption{Warping path in warping matrix}
\label{fig:warpingpath}
\end{figure}

\begin{equation}
	\min_Pd_P(X,Y)
\label{eq:wrappingpath}
\end{equation}
These constraints include monotonicity or ordering that can be expressed as,

\begin{eqnarray}
	p_x(k+1) &\geq& p_x(k) \label{eq:monotonicity} \\ 
	p_y(k+1) &\geq& p_y(k) \nonumber	
\end{eqnarray}
and that the end points of $X$ are mapped to end points of $Y$,

\begin{eqnarray}
	p_x(1) &=& 1 \label{eq:endpoints} \\ 
	p_y(1) &=& 1\nonumber \\ 
	p_x(T) &=& n \nonumber \\ 
	p_y(T) &=& m \nonumber
\end{eqnarray}
Dynamic time warping solves the problem given in Eq. (\ref{eq:wrappingpath}) using dynamic programming. It constructs a warping matrix $W$ as,

\begin{equation}
	W(i, j) = d(i, j) + S 
\end{equation}
where $S$ is called a step pattern and is given as,

\begin{equation}
	S = \min 
	\begin{cases}
	s(i, j-1) \\
	s(i-1, j) \\
	s(i-1, j-1)
	\end{cases}
\end{equation}

\begin{equation}
	s(p, q) = 
	\begin{cases}
	0 \quad \text{if } p=0, q=0 \\
	\infty  \quad \text{if } p=0, q \neq 0 \\
	\infty  \quad \text{if } p \neq 0, q = 0 \\
	w(p, q) \quad \text{otherwise}
	\end{cases}
\end{equation}

\begin{figure}[t]
\centering
\includegraphics[scale=0.35]{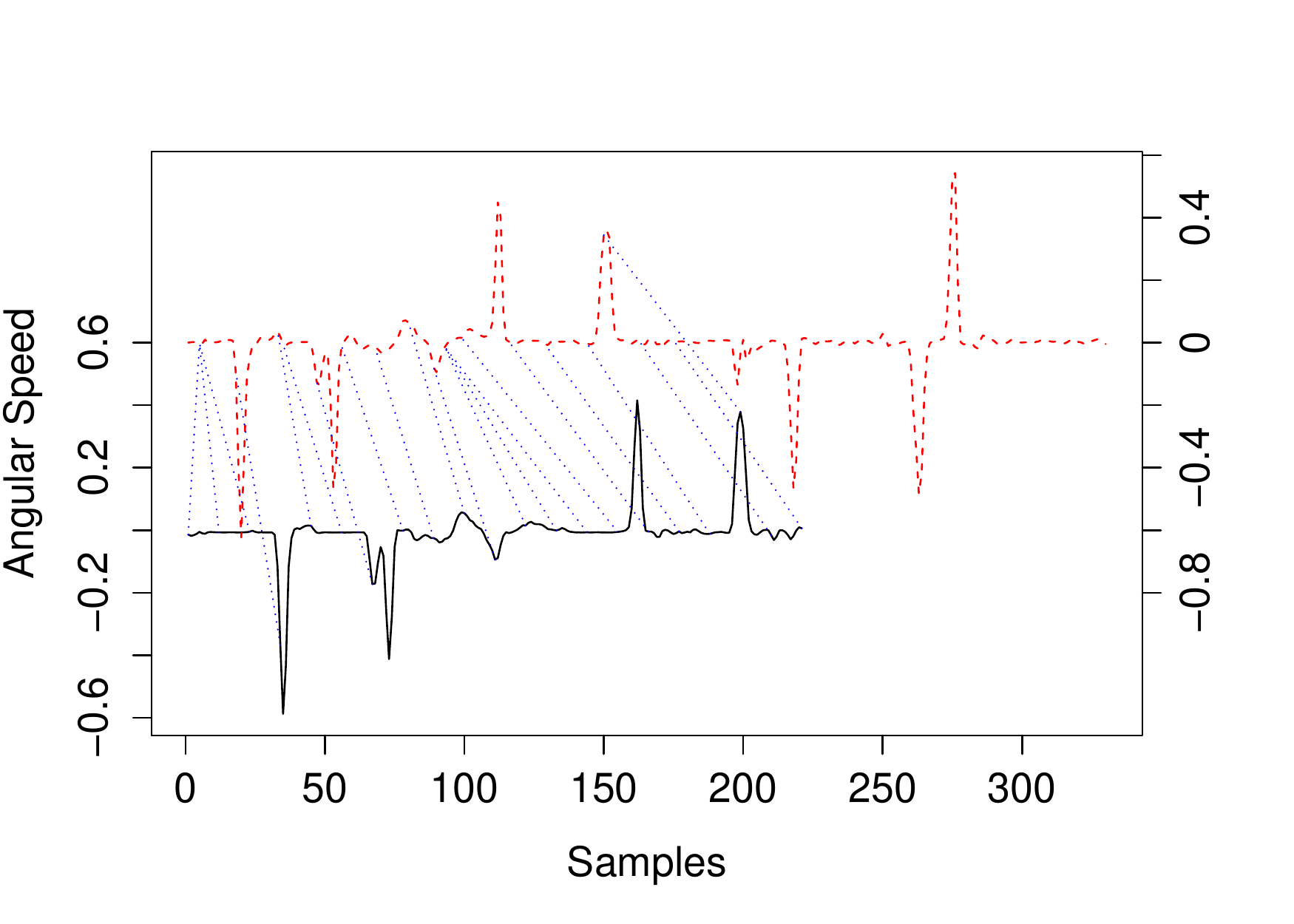}
\caption{Open ended warping}
\label{fig:oedtw}
\end{figure}
Starting from $W(1,1)$, this step pattern constructs a warping path that minimizes the dissimilarity between $X$ and $Y$ while adhering to the constraints given in Eq. (\ref{eq:monotonicity}) and Eq. (\ref{eq:endpoints}). When this recursion terminates, $W(n, m)$ contains the dissimilarity or distance between $X$ and $Y$ which can be normalized according to Eq. (\ref{eq:dtwnormalized}). There are several different types of step patterns~\cite{dtwbook} that can be used in DTW and additional windowing constraints can also be imposed on $P$. However, our experience with our gyroscope dataset indicates that windowing constraints unnecessarily restrict DTW and therefore should not be used for this application. The time complexity of DTW is $\mathcal{O}(nm)$. Fig. (\ref{fig:dtw}) shows a mapping between two sequences computed using DTW. The matched points of two sequences are connected with dotted blue lines. Fig. (\ref{fig:warpingpath}) shows the corresponding wrapping path and the warping matrix where $i$ grows horizontally towards right and $j$ grows vertically up.

\subsection{Open Ended Warping}

\begin{figure}[t]
\centering
\includegraphics[scale=0.4]{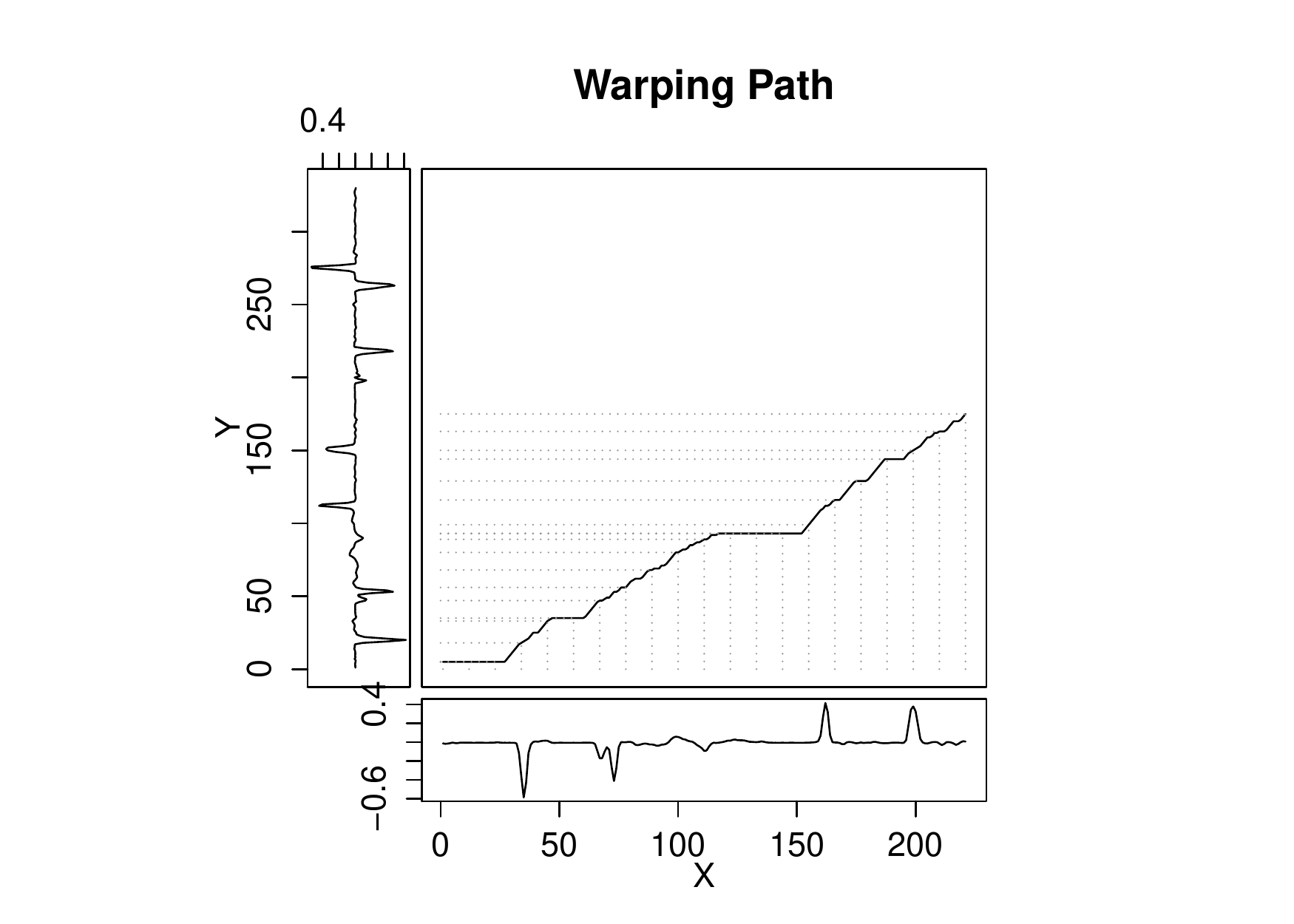}
\caption{Warping path for open ended matching}
\label{fig:oewarpingpath}
\end{figure}

\begin{figure}[t]
\centering
\includegraphics[scale=0.4]{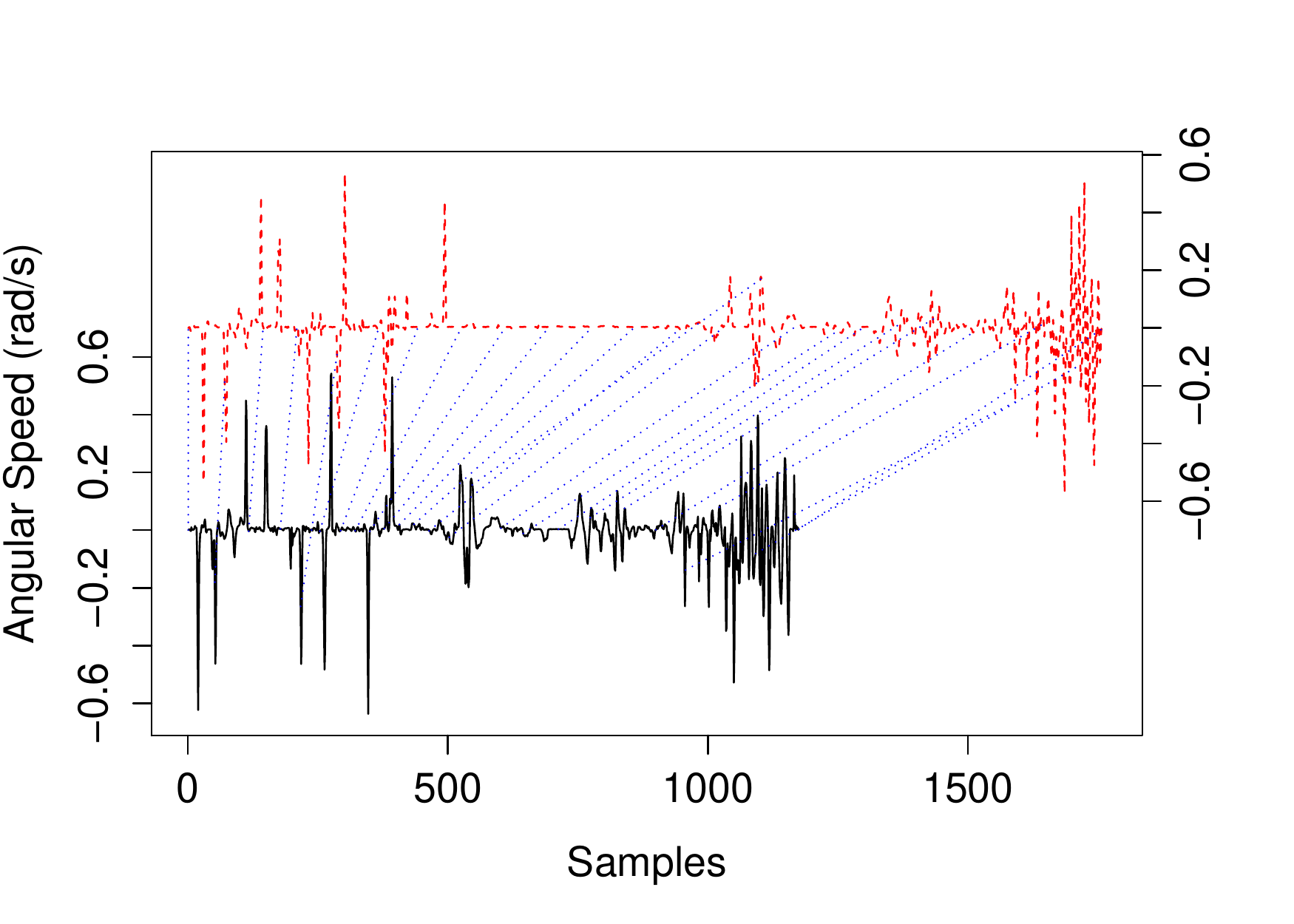}
\caption{DTW mapping for two example journeys}
\label{fig:completejourney}
\end{figure}

In the previous section, we presented dynamic time warping based approach to compare two journeys. While it can be used to compare complete journeys, it does not work when comparing a partial journey with a complete previous journey. For example, if we want to compare the output of the gyroscope captured up to the current moment in the journey to a previous journey to predict the destination and remaining route, we cannot use the basic DTW approach because it maps every point between the two sequences $X$ and $Y$ and forces end points of $X$ to map to those of $Y$. We, therefore, need a means of comparing a given sequence $X$ of gyroscope output to all possible truncated versions $Y^j (j=1,2, \hdots, m)$ of $Y$, gyroscope output from a previous journey, and choose the best possible match i.e.,

\begin{equation}
	\min_j d(X, Y^j)
\end{equation}
We can solve this problem by executing the basic DTW as described in the previous section. And then taking the smallest cumulative distance in $W(n, j)$ column of the warping matrix instead of $W(n, m)$. This gives us the prefix length $j$, the distance between $X$ and $Y^j$ and the best matching prefix $Y^j$. This is known as open-ended DTW~\cite{oedtw}. It is also possible to match $X$ with all possible sub-sequences $Y^{i, j} (i=1, 2, \hdots, m), j>i$ of $Y$,
 
 \begin{equation}
	\min_{i,j} d(X, Y^{i,j})
\end{equation}
This is known as open-beginning open-ended DTW~\cite{streammonitoring} and requires only one additional $n \times m$ matrix to keep track of the starting index of sequences. The time complexity of both open-ended and open-beginning open-ended DTW is $\mathcal{O}(nm)$. Fig. (\ref{fig:oedtw}) shows the open-beginning open-ended DTW of two time series of gyroscope data. It shows that it is able to correctly identify the matching portions of the two sequences. Sometimes trips start or end in car parks where turns vary from one trip to another as a user parks in different locations. Open-beginning open-ended DTW is useful in ignoring such turns at the start or end and matching only the intermediate part of the trip. Fig. (\ref{fig:oewarpingpath}) shows the corresponding warping path and the warping matrix where $i$ grows horizontally towards right and $j$ grows vertically up. Fig. (\ref{fig:completejourney}) shows the DTW mapping between the two complete journeys that we initially discussed in Section \ref{sec:approachoverview}.

\subsection{Route Mining}
\label{sec:route_mining}

We have described our time warping based approach that is able to distinguish between two routes and is not affected by the type, orientation or position of the sensing device inside the vehicle: we will show results of the evaluation of these aspects in the next sections. We now discuss how we can mine the axis aligned angular speeds captured by the gyroscope to identify important or significant routes that a user frequently travels on. 

Let us suppose that during the bootstrap period when a user starts using the system, it captures and aligns the angular speeds of the vehicle from $n$ journeys made the by user. We can then compute normalized DTW distances between each pair of journeys to construct an $n \times n$ dissimilarity matrix $D$. This dissimilarity matrix $D$ can then be used with any of the several clustering algorithms~\cite{clustering} to group similar journeys or routes together. Clusters with large number of members then represent routes frequently used by the user.
In the next few sections we will show how this method performs over a set of collected real traces.

\section{Accuracy Evaluation}
In this section we report the results of the evaluation we conducted to study the feasibility of the approach.
We first show how the phone position and type of handset are not effecting the performance. We then show how routes collected through phones handed out to participants can be clustered automatically into significant routes.
\subsection{Phone Position and Handset Independence}

\begin{figure}[t]
\centering
\includegraphics[scale=0.22]{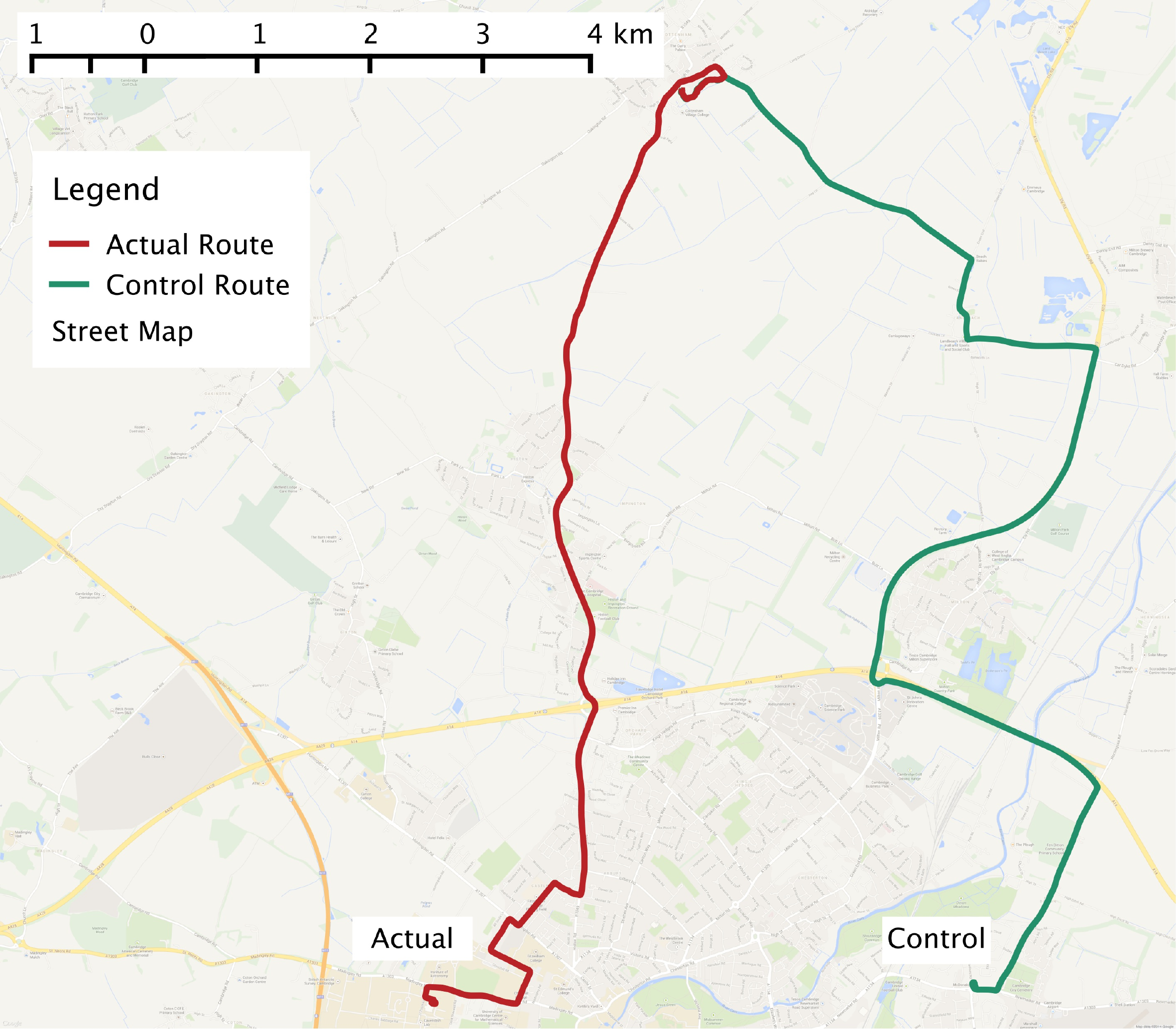}
\caption{Trajectories of actual and control paths}
\label{fig:paths}
\end{figure}

\begin{figure*}[bt]
\centering
\begin{minipage}{0.3\linewidth}
  \includegraphics[scale=0.32]{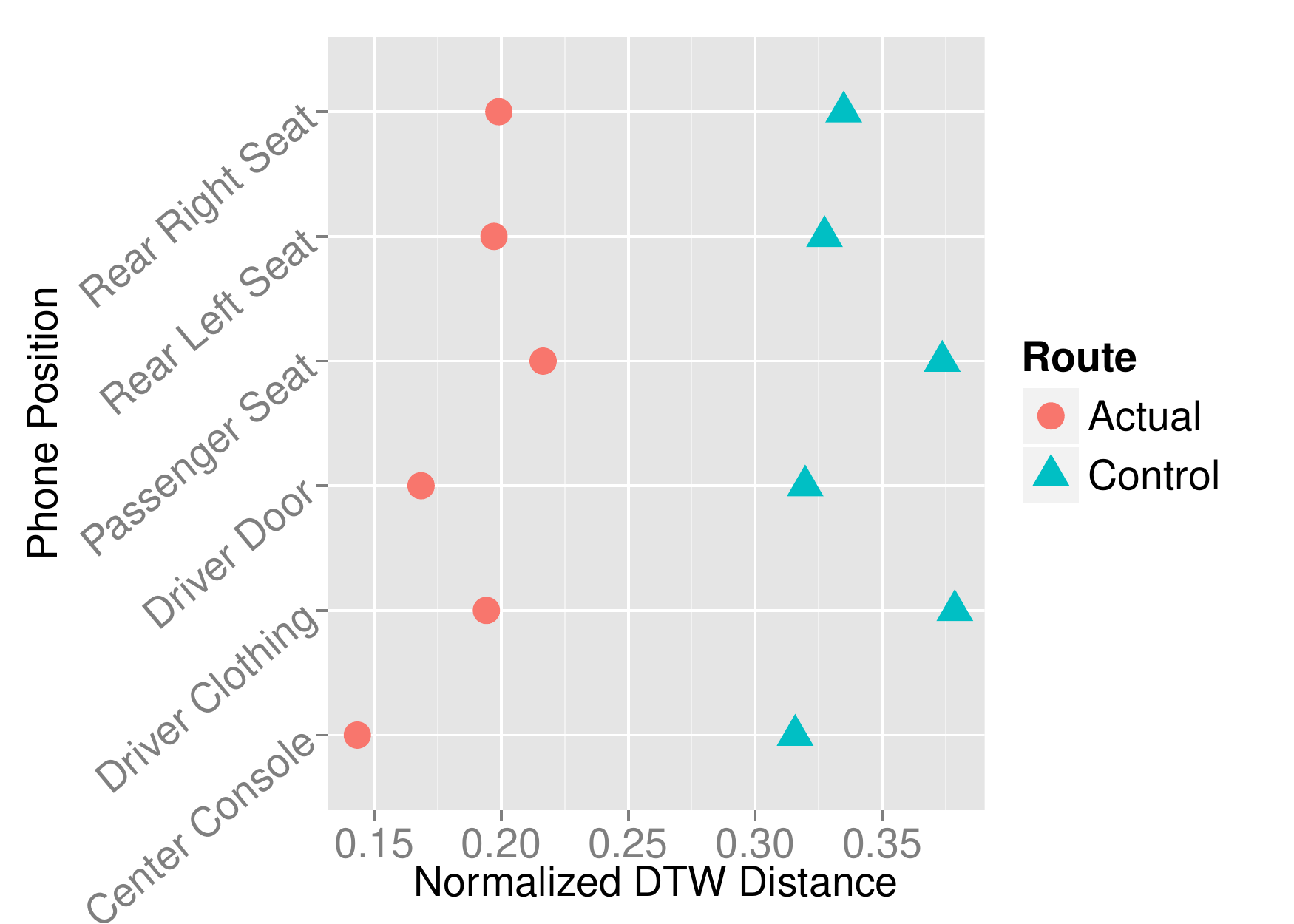}
  \caption{Phone position}
  \label{fig:phonelocation}
\end{minipage}
\begin{minipage}{0.3\linewidth}
  \includegraphics[scale=0.32]{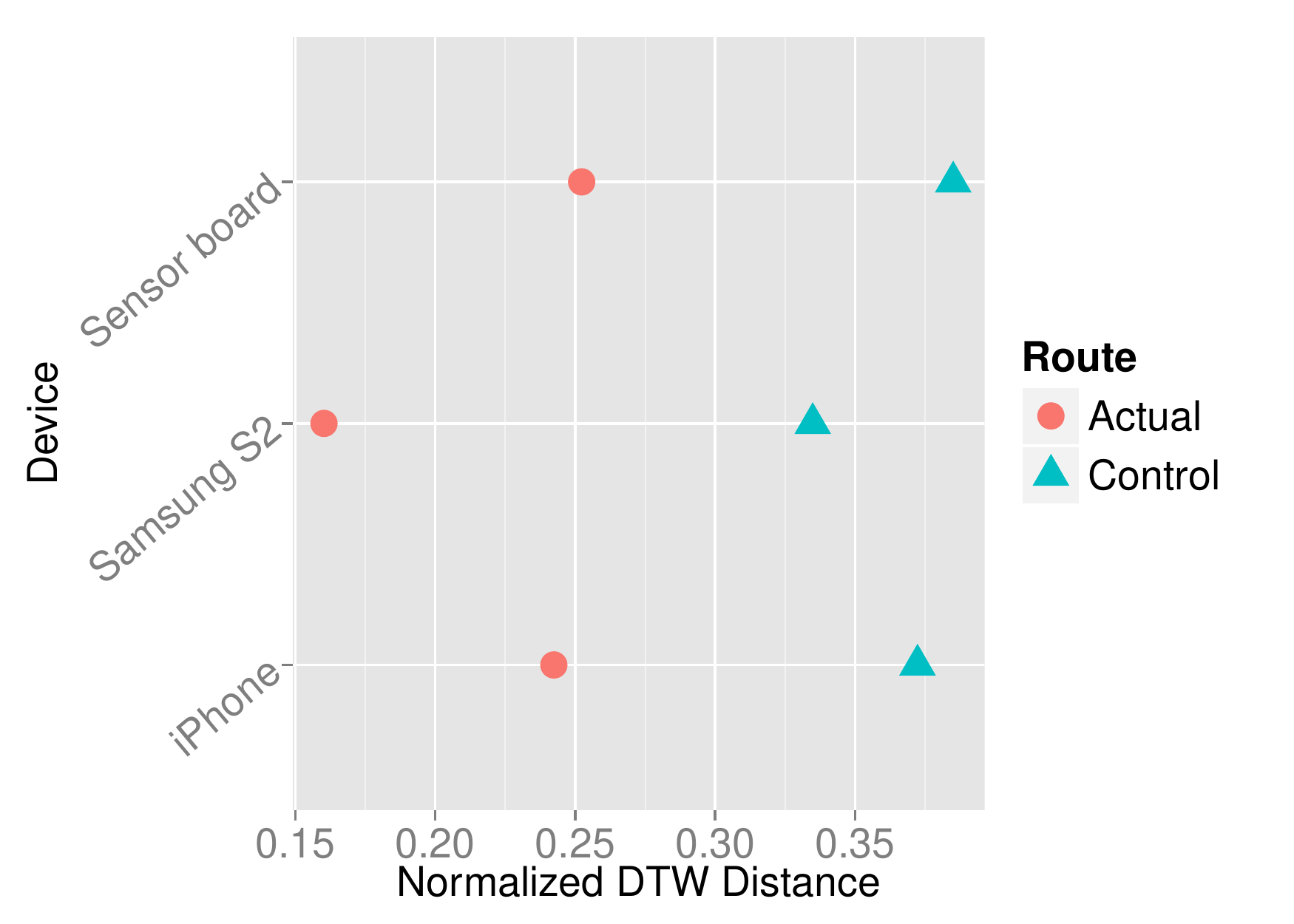}
  \caption{Different devices}
  \label{fig:devicetype}
\end{minipage}
\begin{minipage}{0.3\linewidth}
  \includegraphics[scale=0.32]{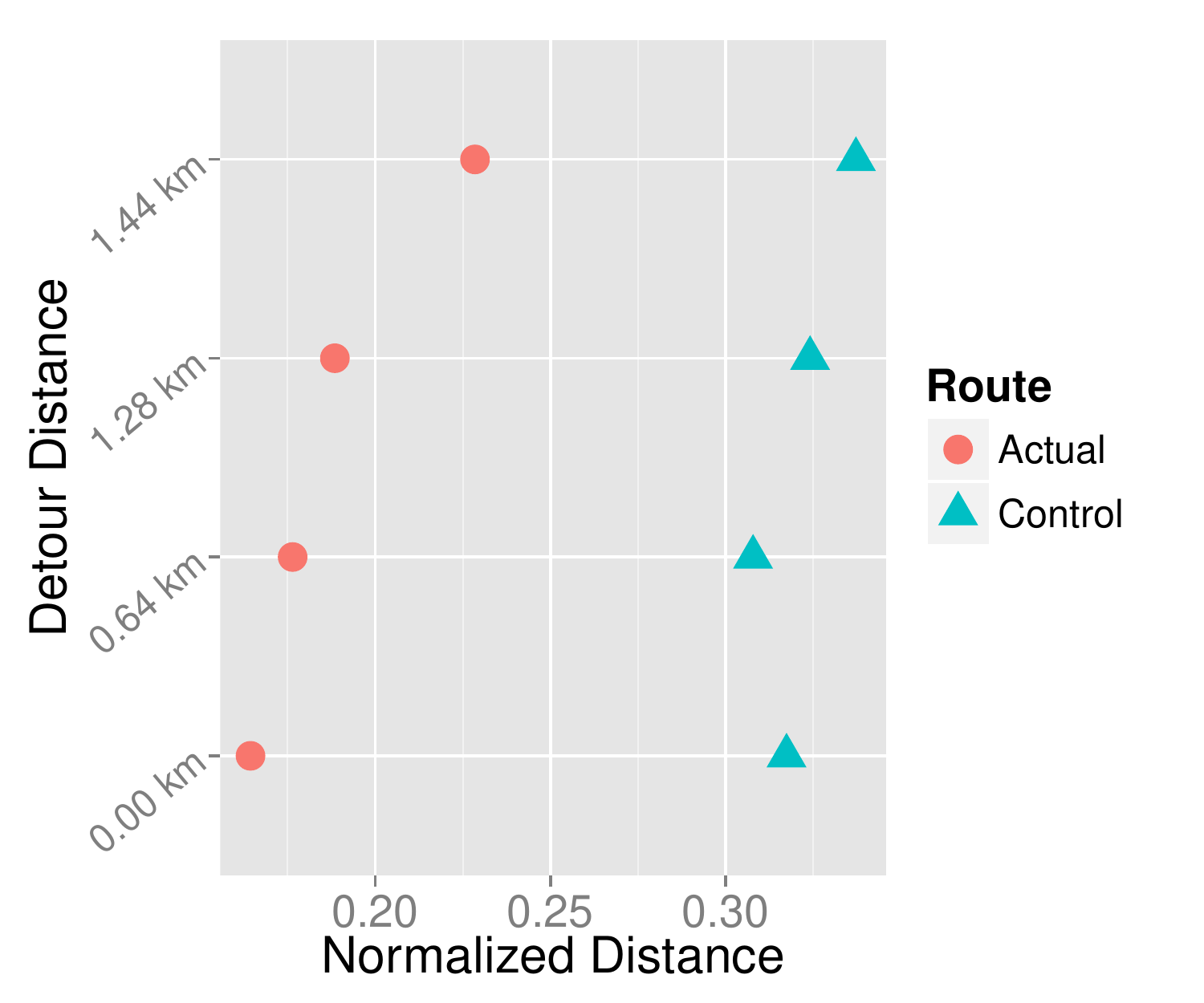}
  \caption{Small detours}
  \label{fig:detours}
\end{minipage}
\end{figure*}

In order to demonstrate that our route matching approach is not affected by  phone position inside the traveling vehicle, we conduct a small experiment. We collect a trace of angular speed from a phone inside the vehicle on a typical commute route from source A to destination B. The route is about $12$ km long and takes about $20$ minutes to complete in ordinary driving conditions. We will refer to this as \emph{actual} route. We also collected  a trace from a second route of similar distance from source A to another destination C. We will refer to this as \emph{control} route. Fig. (\ref{fig:paths}) shows both of these routes on a street map. For both of these cases, we placed the phone in the central console between the driver and the front passenger seat. Over the course of next few days, we collected several test traces on the commute route from A to B by placing the phone at various positions inside the vehicle in an arbitrary orientation. We then compared these test traces with the initial data collected from both the actual and control route using dynamic time warping. Fig. (\ref{fig:phonelocation}) shows the normalized DTW distances calculated between the test cases and the original data from both the actual and control routes. A red circle represents normalized distance calculated between the test trace for a particular phone position in the vehicle and the initial data from the actual route. A blue triangle represents the distance between the same test trace and the initial data from the control route. It shows that under the normalized DTW distance, the test cases are consistently closer to the actual route as compared to the control route irrespective of the phone position and orientation in the vehicle. It is, therefore, possible to correctly recognize the actual route with the phone is located anywhere in the vehicle in an arbitrary orientation. We also used different devices to capture test traces from the actual route and compared these in a similar manner. Fig. (\ref{fig:devicetype}) shows the normalized distances between the test traces from a Samsung S2, an iPhone 5s and an embedded sensor board (described later) and the initial data from the actual and control route captured with a Samsung S2. It shows that test traces are closer to the actual route under the normalized DTW distance. This demonstrates that our approach works across devices as well. We also evaluate cases where small detours are taken from the actual route while traveling from source A to destination B. Fig. (\ref{fig:detours}) shows that even in the presence of these small detours, the traces taken from actual route are close to each other under DTW distance as compare to that from the control route. The DTW distance from the actual route gradually increases with the length of the detour.

\subsection{Route Clustering}

\begin{figure}[t]
\centering
\includegraphics[scale=0.35]{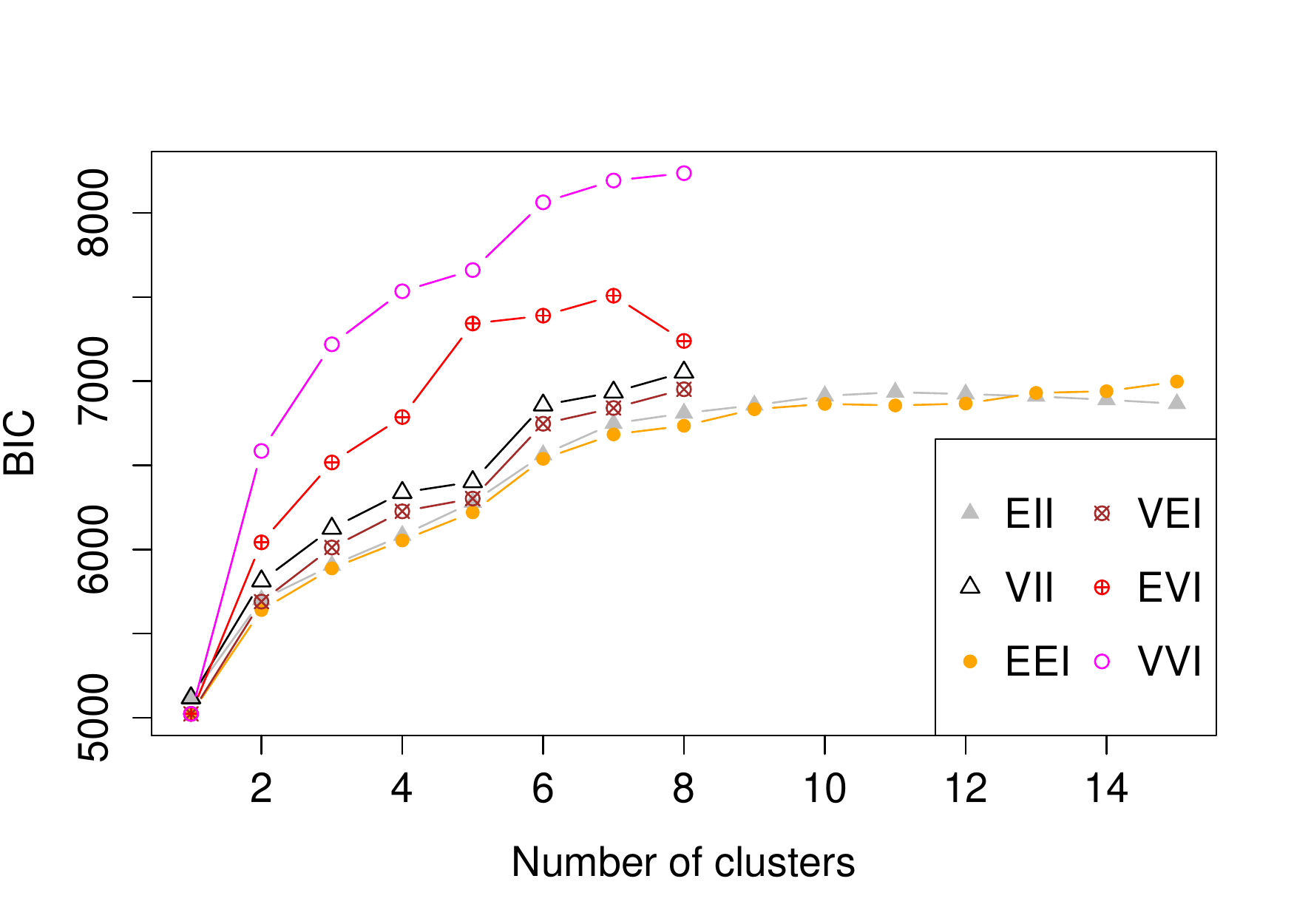}
\caption{Estimated number of clusters}
\label{fig:bic_gyro_cluster}
\end{figure}

In order to demonstrate the feasibility of route mining, we collected a labelled dataset involving several typical journeys including commuting, leisure and other traveling undertaken over a course of two months in two different cities. We captured over $240$Mbytes of sensor data from about $400$ kilometers of driving under varying traffic conditions including urban, suburban and highway driving. This includes normal daily driving with lane changes, varying turning speeds and overtaking other vehicles. It also includes journeys that start or stop in car parks where turns vary between trips. For each journey, we collected accelerometer, gyroscope and GPS sensor data and recorded the source and destination of the journey. The dataset includes a total of $43$ journeys over $15$ routes (source destination pairs). The shortest route is about $4$km where as the longest route is just over $90$km long. Some parts of certain routes also overlap for different source or destination locations. For example, when routes originating at different source locations end in the same destination and when routes start at the same source but end at different destination locations.

We use k-medoids~\cite{pam} to partition this dataset into clusters. It is a robust version of the popular k-means clustering algorithm and uses actual data points as cluster centers which is useful in our application because we intend to compare new routes with these clusters. However, any of the several clustering algorithms available in literature can be used for this partitioning. A system based threshold $\tau$ can be defined where any cluster with more than $\tau$ members is marked as a significant route. Thus the user will have to make at least $\tau$ journeys on a route for it to be marked as a significant route of the user. Before we can cluster the routes with k-medoids, we have to determine k, the number of clusters in our dataset. One possible approach to this is to use finite mixture models~\cite{mixturemodels} that view the data as originating from different processes each modeled as a Gaussian distribution. The number of clusters can then be estimated by varying the number of processes and model parameters and observing the Bayesian Information Criterion (BIC)~\cite{bic} for model selection. Fig. (\ref{fig:bic_gyro_cluster}) shows that BIC is maximized when the data is modeled as originating from 8 different processes among various model parametrizations. This suggests that there are 8 clusters in the dataset.

\begin{figure}[t]
\centering
\includegraphics[scale=0.45]{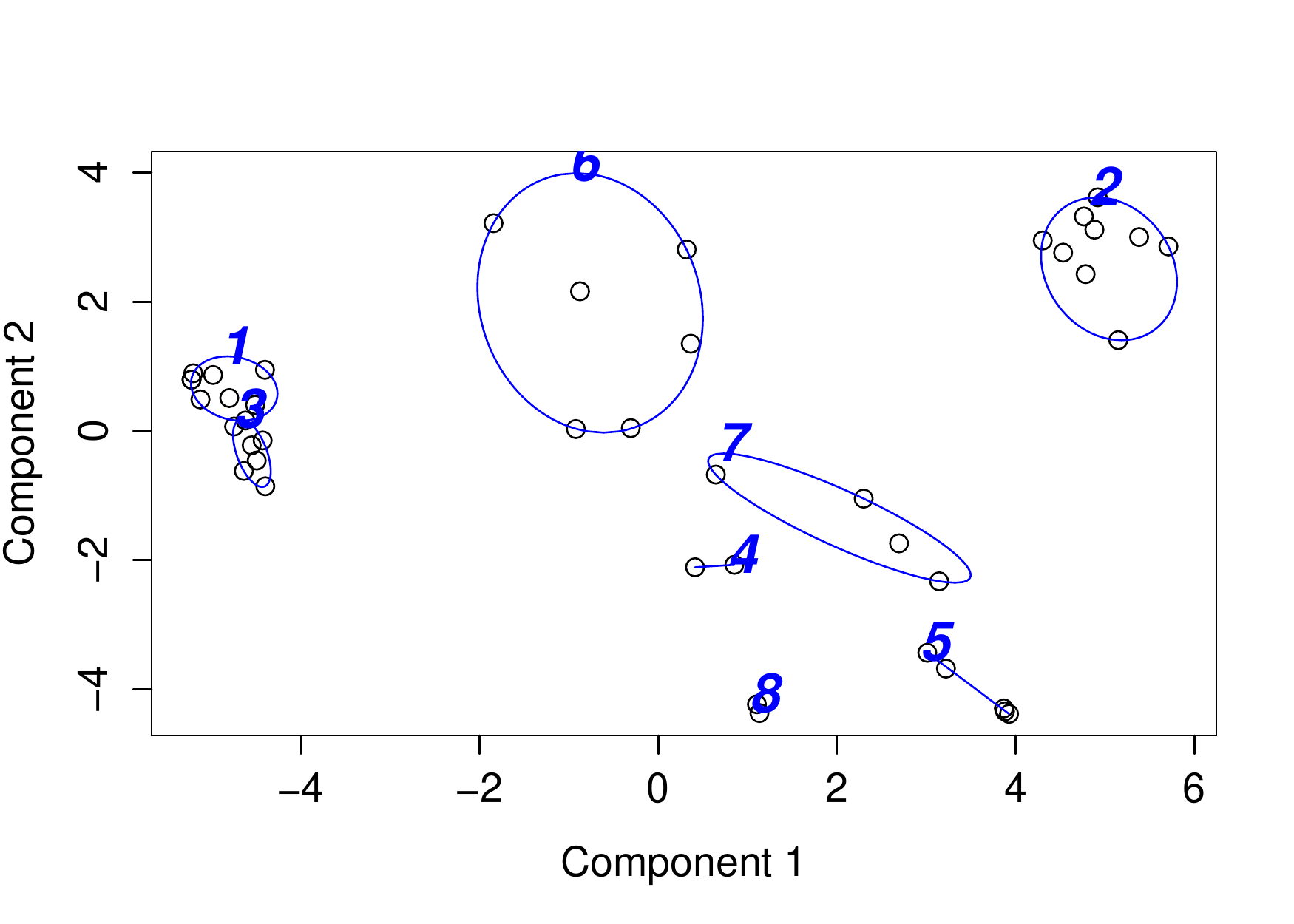}
\caption{Clustering of routes based on angular speed}
\label{fig:gyro_cluster}
\end{figure}

Fig. (\ref{fig:gyro_cluster}) shows the partitioning clusters generated by k-mediods with $k=8$ from the dissimilarity matrix $D$ of normalized DTW distances between the time series of angular speeds from different journeys. It shows that the routes are organized in dense well separated clusters. Similar routes are clustered together into a single cluster or in nearby clusters. For example, cluster 1 contains all the journeys from source A to destination B and the journeys in reverse direction are grouped in the adjacent cluster 3. Three of the four routes grouped together in cluster 7 originate from one source whereas the fourth route has a different starting location. However, all four routes have a common destination and therefore overlap with each other.

To evaluate this clustering more objectively, we compare it to GPS trajectories of these journeys. As the journeys can be of unequal length and partially overlapping, we have to create a mapping between each pair of trajectories  using location data. DTW has been used for computing similarities between $2$D trajectories~\cite{dtwtrajectory1} and GPS data~\cite{dtwtrajectory2} in trajectory analysis literature. When used with $2$D (or $3$D) location data, it creates a mapping between each location point $p_1$ in trajectory $T_1$ and location point $p_2$ in the second trajectory $T_2$ and then uses the supplied distance function (great circle distance for latitude longitude coordinates) between mapped points in $T_1$ and $T_2$ to calculate similarity between $T_1$ and $T_2$. This gives us the intended output i.e. the larger the overlap between two trajectories, the smaller the dissimilarity and vice versa. 

\begin{figure}[t]
\centering
\includegraphics[scale=0.45]{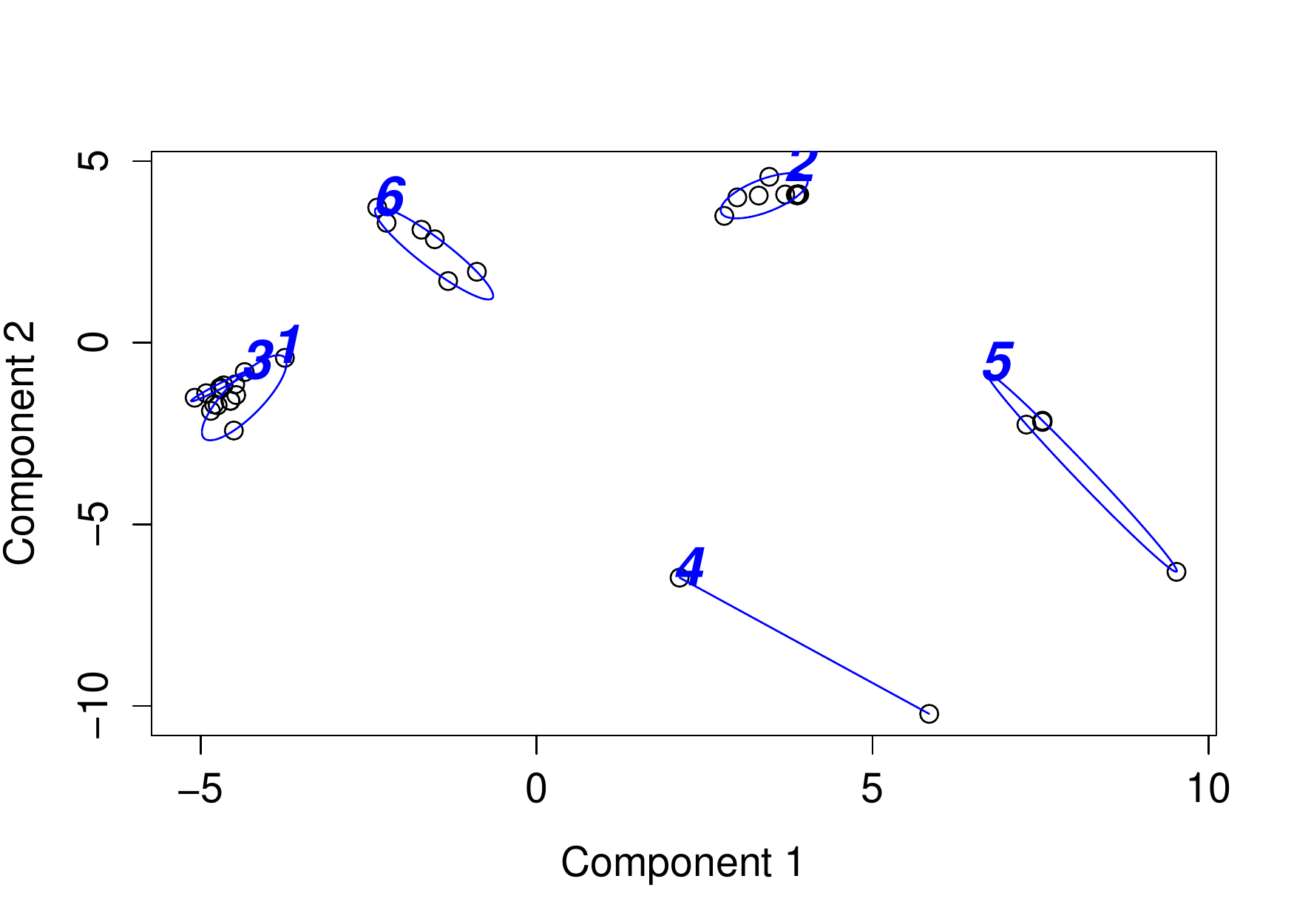}
\caption{Clustering of routes based on GPS trajectories in city A}
\label{fig:gpscam_cluster}
\end{figure}

\begin{figure}[t]
\centering
\includegraphics[scale=0.45]{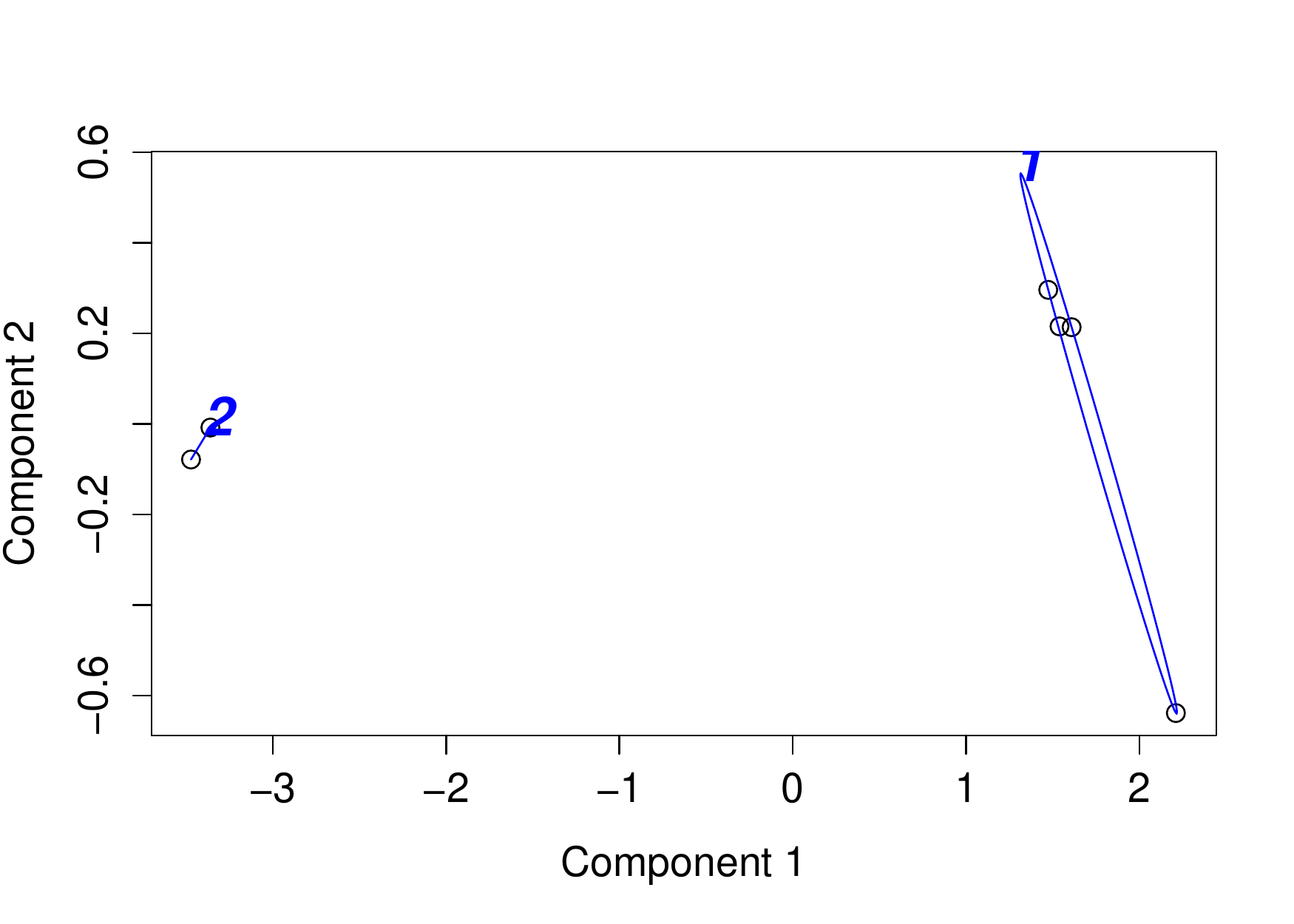}
\caption{Clustering of routes based on GPS trajectories in city B}
\label{fig:gpsips_cluster}
\end{figure}

As we use geographic distances and the fact that the journeys took place in two different cities, initial clustering of GPS trajectories leads to two clusters, one in each city. We then perform a second round of clustering on routes in each of the two cities individually. We again use the same approach that we used for angular speed clustering. We use BIC to estimate the number of clusters in each city and then use k-medoids to perform clustering. This leads to 6 clusters in city A and 2 clusters in city B. These are shown in Fig. (\ref{fig:gpscam_cluster}) and Fig. (\ref{fig:gpsips_cluster}) respectively. While the shapes and separation between clusters of GPS trajectories are different from those based on angular speeds, the figures show that the overall partitioning is very similar. Clusters $\lbrace1, 2, \hdots, 6\rbrace$ of Fig. (\ref{fig:gpscam_cluster}) correspond to the clusters with similar labels in Fig. (\ref{fig:gyro_cluster}) and clusters $1$ and $2$ of Fig. (\ref{fig:gpsips_cluster}) correspond to clusters $7$ and $8$ of Fig. (\ref{fig:gyro_cluster}). A corrected rand index~\cite{randindex} and variation of information distance (VI)~\cite{vidist} are two measures that are extensively used to compare two different clustering outputs to see how similar or different they are. We compute both of these measures to compare the clusters based on angular speeds to those based on GPS trajectories. A rand index of $r=1$ and variation of information distance $\text{VI}=0$ between the two clusterings show that these clusterings are identical i.e. individual journeys or routes that are clustered together based on GPS trajectories are in the same clusters when partitioning is performed with gyroscope angular speeds. This shows that if we want to sense and later recognize routes or journeys, angular speeds from the gyroscope can provide us with the same information as the GPS sensor.

\subsection{Grid Road Networks}

\begin{table}[t]
\centering
\caption{Rules for encoding turning angles into alphabets}
\label{tab:turnEncoding}
\begin{tabular}{|c|l|c|}
\hline
\textbf{\begin{tabular}[c]{@{}c@{}}Encoding \\ Rule\end{tabular}} & \multicolumn{1}{c|}{\textbf{\begin{tabular}[c]{@{}c@{}}Turn \\ Information\end{tabular}}} & \textbf{Encoded Alphabet} \\ \hline
$-30\degree \le t < -15\degree$ & Slight Right              & S                         \\ \hline
$t < -30\degree$               & Right                     & R                         \\ \hline
$15\degree < t \le 30\degree$   & Slight Left               & T                         \\ \hline
$t > 30\degree$             & Left                      & L                         \\ \hline
$-15\degree \le t \le 15\degree$  & Ignore                    & N/A                       \\ \hline
\end{tabular}
\end{table}

In the last few subsections, we showed that it is possible to recognize a trip by inspecting variations of angular speeds experienced by the phone resulting from turns taken by the vehicle during the journey. But is it possible to use this approach in urban areas with Manhattan style grid based road network where turns at majority of the intersections are similar?  We argue that on a grid type road network, a reasonably long route can be uniquely identified because it exhibits a unique \emph{sequence of turns} even if individual turns are similar. 

\begin{figure}[t]
\centering
\includegraphics[scale=0.24]{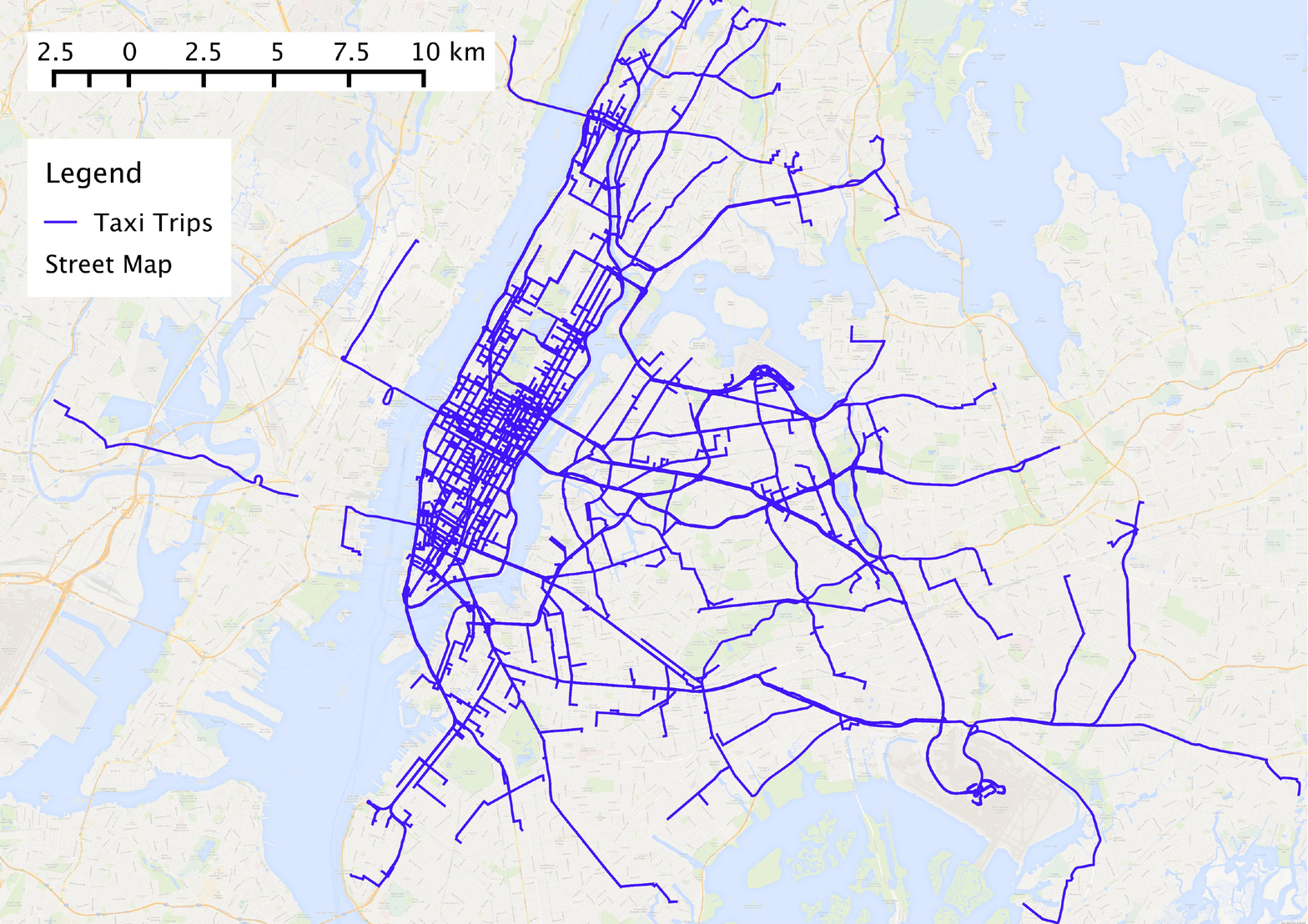}
\caption{Taxi trips used for calculating route similarity in Manhattan New York}
\label{fig:taxiroutes}
\end{figure}

\begin{figure}[t]
\centering
\includegraphics[scale=0.35]{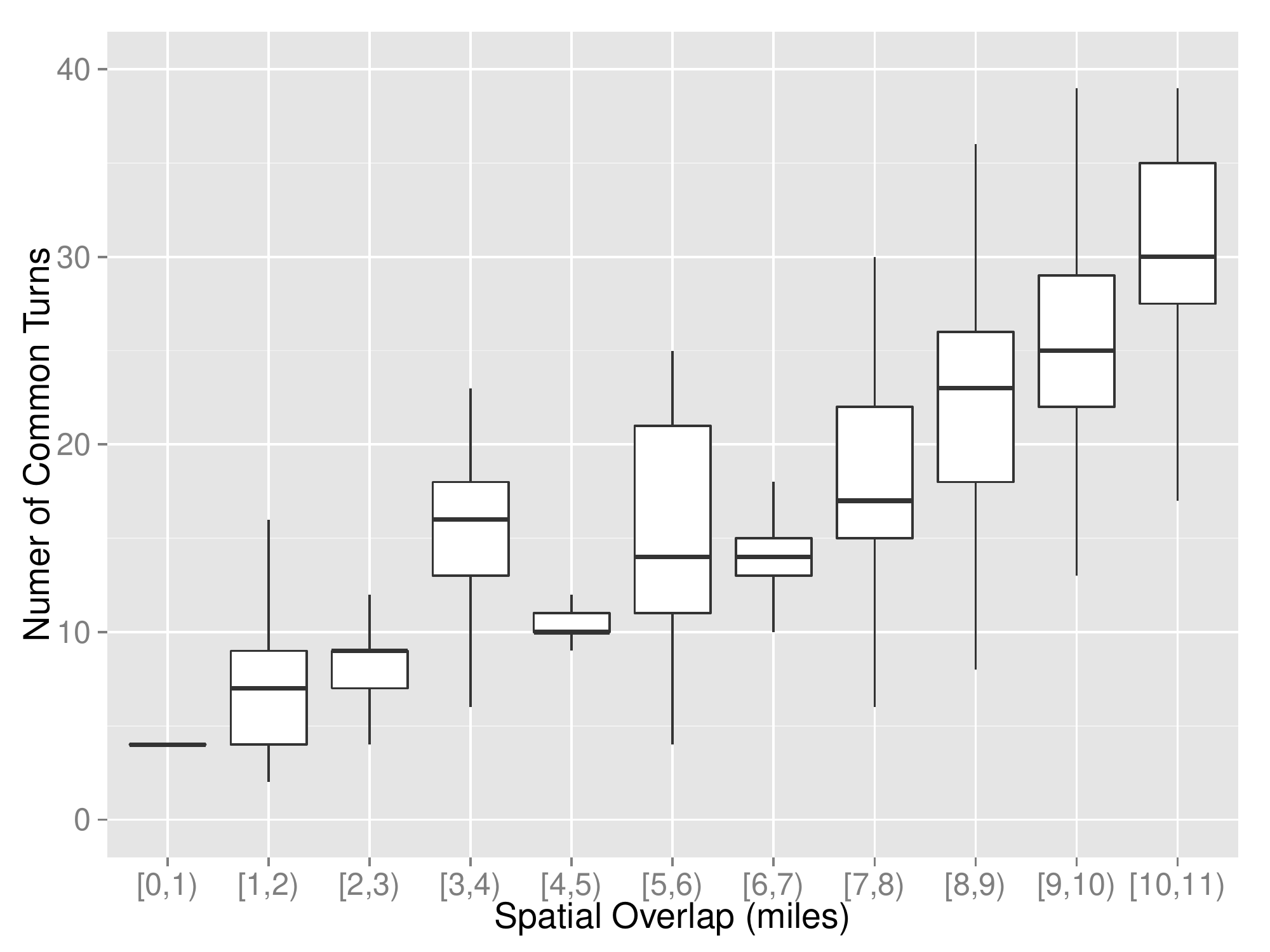}
\caption{Turn based route similarity in Manhattan}
\label{fig:tripsim}
\end{figure}

In order to show this, we use a dataset recently released by the New York metropolitan transportation authority containing information of real taxi trips in New York\footnote{Taxi Trips \url{http://chriswhong.com/open-data/foil_nyc_taxi/}}. For each trip, the dataset contains travelled distance, pick up and drop locations and other metadata. According to 2009 National Household Travel Survey~\cite{travelsurvey}, the average length of a typical vehicle trip in a metropolitan area in US is $8.8$ miles. We, therefore, take a random sample of $1000$ taxi trips in New York from the dataset with an average distance of $9$ miles. As the dataset does not contain actual routes taken by taxis, we use the A* routing algorithm~\cite{astar} on Open Street Map road network of New York to find road segments that form the shortest route between pickup and drop off locations for each trip. Given these shortest route road segments for each trip, we then compute turning angles for all turns along the computed route, encode these turns into alphabets according to the rules given in Table (\ref{tab:turnEncoding}) and concatenate resulting characters into a string in the same order in which associated turns occur along the route. Thus for each trip, we have a trajectory in the form of road segments and a string with characters representing turns along this trajectory. We will refer to this as a maneuver string. Fig. (\ref{fig:taxiroutes}) overlays all of these trajectories on a single layer showing that these trips span most of Manhattan and the surrounding urban area. In order to determine similarity between two trips based only on turns, we find the longest common sequence of characters between maneuver strings for each pair of trips. We then compare it to the spatial overlap between trajectories of same pair of trips. Fig. (\ref{fig:tripsim}) plots the number of turns in the longest common sequence of turns against the spatial overlap for all possible combinations of $1000$ taxi trips. It shows that for trajectories with no or little overlap (in the [0,1) miles interval), the number of turns in the longest common sequence of turns on average is just $4$. Whereas the average number of turns in a typical trip trajectory in our sample is $22$. This shows that trips of average length can be uniquely identified based solely on turn information even on grid type road networks.  

\section{Energy Evaluation}

\begin{figure}[t]
\centering
\includegraphics[scale=0.37]{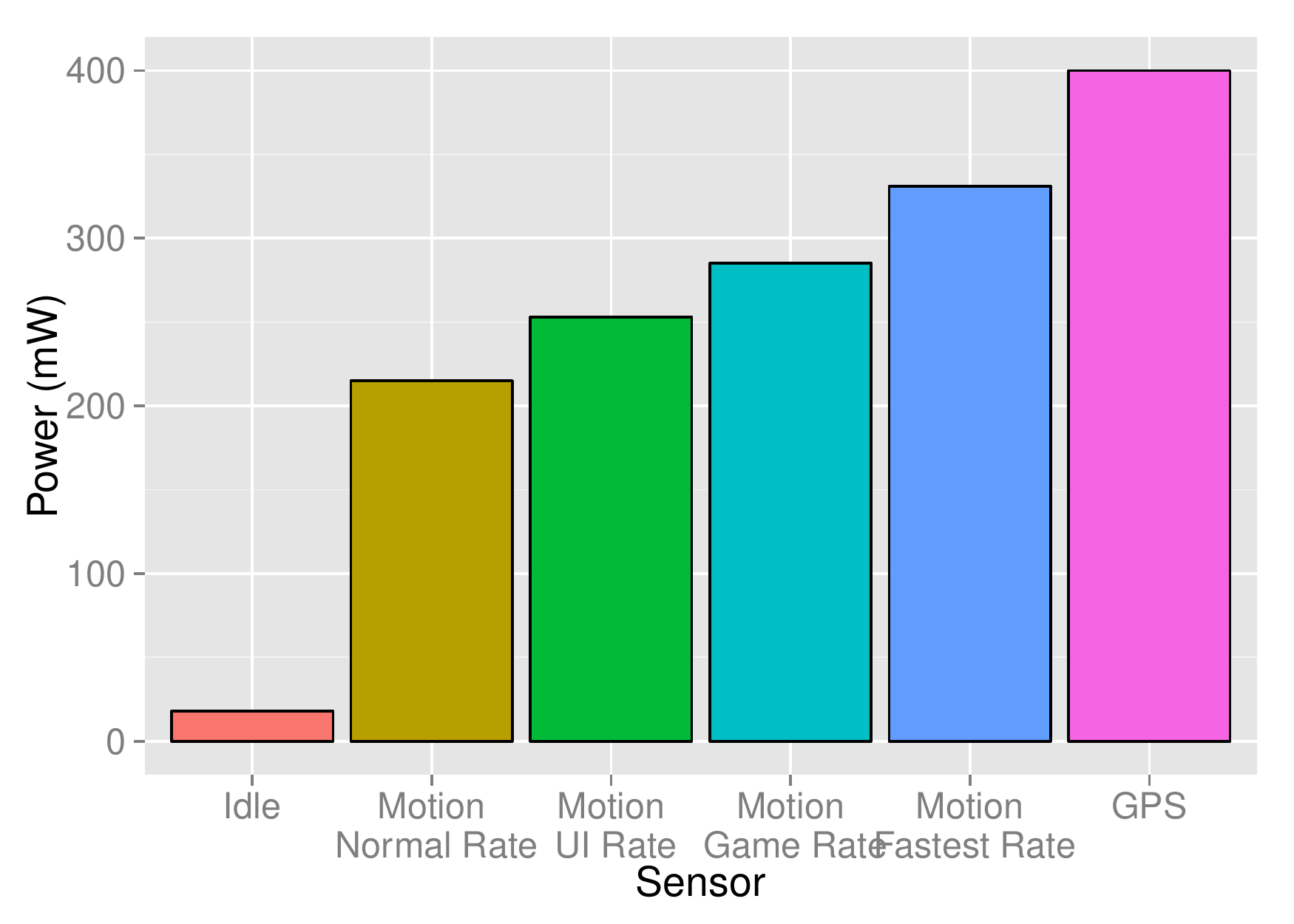}
\caption{Energy consumption of motion sensors and GPS on Samsung S2 smartphone}
\label{fig:motionenergy}
\end{figure}

In the last few sections, we showed that a smartphone's accelerometer and a gyroscope can be accurately used to sense and recognize routes frequently used by a driver. We now discuss first our implementation of the system on a smartphone and then our co-processor based implementation to then discuss their performance.

\subsection{Smartphone Based System Evaluation}
We start by investigating the energy consumption of different sensors on a smartphone. We use a Monsoon power monitor\footnote{\url{http://www.msoon.com/LabEquipment/PowerMonitor/}} to measure energy consumption of a Samsung S2 smartphone running Android OS while sampling the accelerometer and the gyroscope. The Android API defines four different sampling rates (Normal, UI, Game and Fastest) that an application developer can specify to sample from these sensors. We measure the energy consumption of the phone with each of these sampling rates. We also measure the energy consumption when the phone is in idle state and when it is sampling the GPS sensor. Fig. (\ref{fig:motionenergy}) compares the energy consumption of all of these states. It shows that even with the lowest sampling rate, the energy consumption of sampling the accelerometer and the gyroscope is slightly more than $50\%$ of the energy consumed with the GPS sensor. Increasing the sampling rate to fastest leads to energy consumption that is more than $80\%$ of that with the GPS sensor. These measurements were taken with the phone only sampling the sensors and no additional post processing or computations were performed on the sensor data. However, the sensor data from MEMS accelerometers and gyroscopes usually contains high frequency noise that requires filtering and post processing before it can be used in the application. This processing will lead to even higher energy consumption. This shows that using these sensors for sensing might not be an energy efficient option when compared to GPS. 

Several software based duty cycling approaches have been proposed~\cite{jigsaw, rateadaptive} to address high energy consumption of sensing on smartphones, however, these still do not lead to significant energy savings due to high energy consumption of the main processor of a smartphone~\cite{littlerock}. A more effective approach is to use a sensing system with its own dedicated low power processor. This approach has been recently adopted by device manufacturers and co-processors have been integrated in modern smartphones like Apple's iPhone 5s\footnote{\url{https://www.apple.com/iphone-5s/specs}} and Motorola's Moto X\footnote{\url{http://www.motorola.com/us/FLEXR1-1/moto-x-specifications.html}}. Apple's iPhone 5s continuously tracks users activities including walking, running and detecting that the user is in a motorized transport using motion sensors. Motorola's Moto X performs continuous voice sensing and recognition. However, these systems are closed and do not offer developers the ability to program or leverage these co-processors for their own applications. In order to demonstrate the feasibility and energy efficient of our approach on such dedicated co-processors, we use a separate sensing board that includes the necessary sensors and a low power processor and can be connected to a smartphone either through bluetooth or USB interface. However, our approach can be integrated into devices with internal co-processors if the manufacturers open up these systems. This could lead to say an iPhone not only detecting that a user is in a motorized transport but also recognizing that the user is traveling on his or her usual commute route using only the motion sensors. This system and its performance are described next.

\subsection{Co-processor Based System Evaluation}
\label{sec:system_components}
\begin{figure*}[t]
\centering
\begin{minipage}[b]{0.25\linewidth}
  \includegraphics[scale=0.18]{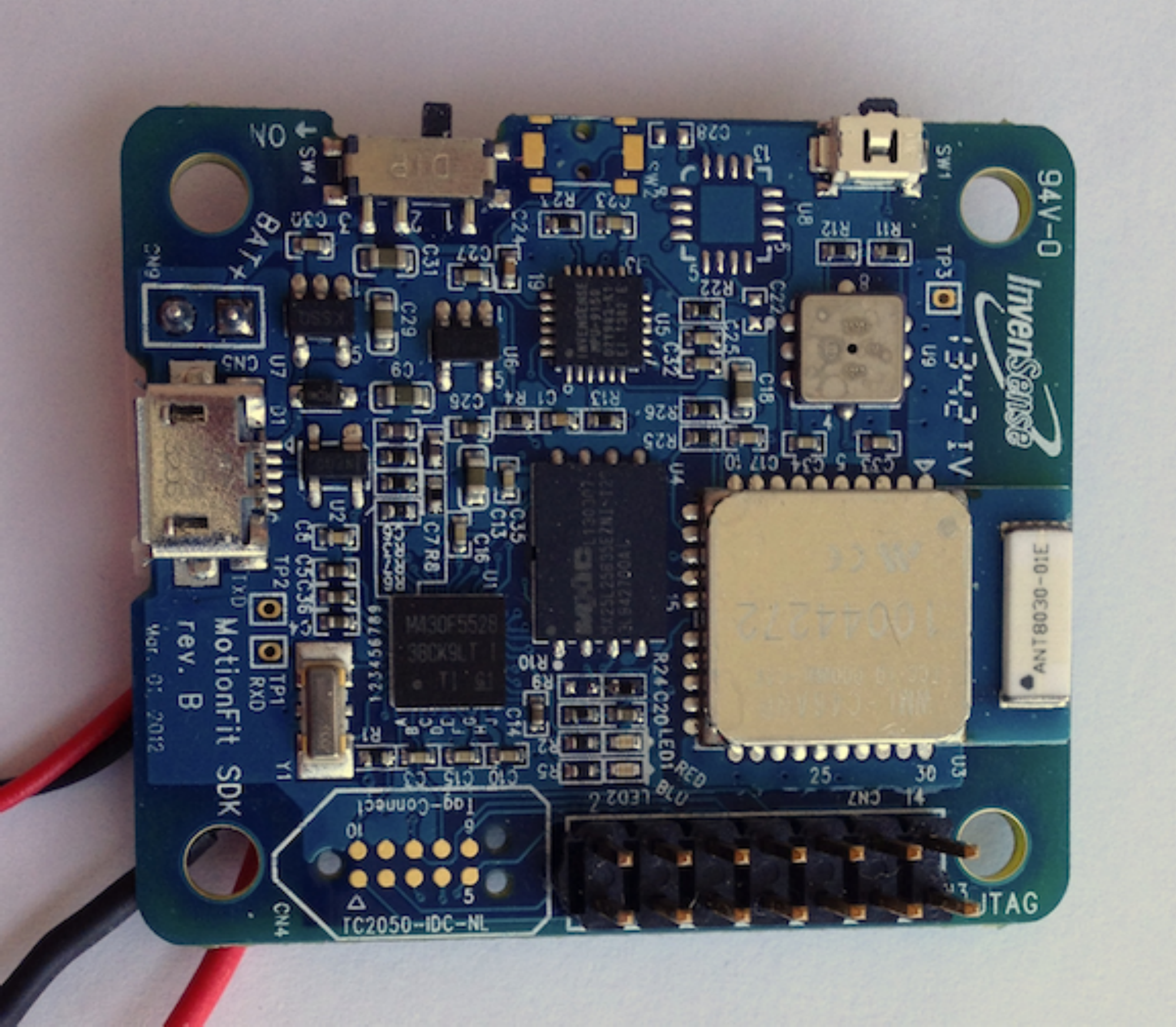}
  \caption{Sensor board}
  \label{fig:sensorboard}
\end{minipage}
\begin{minipage}[b]{0.3\linewidth}
  \includegraphics[scale=0.2]{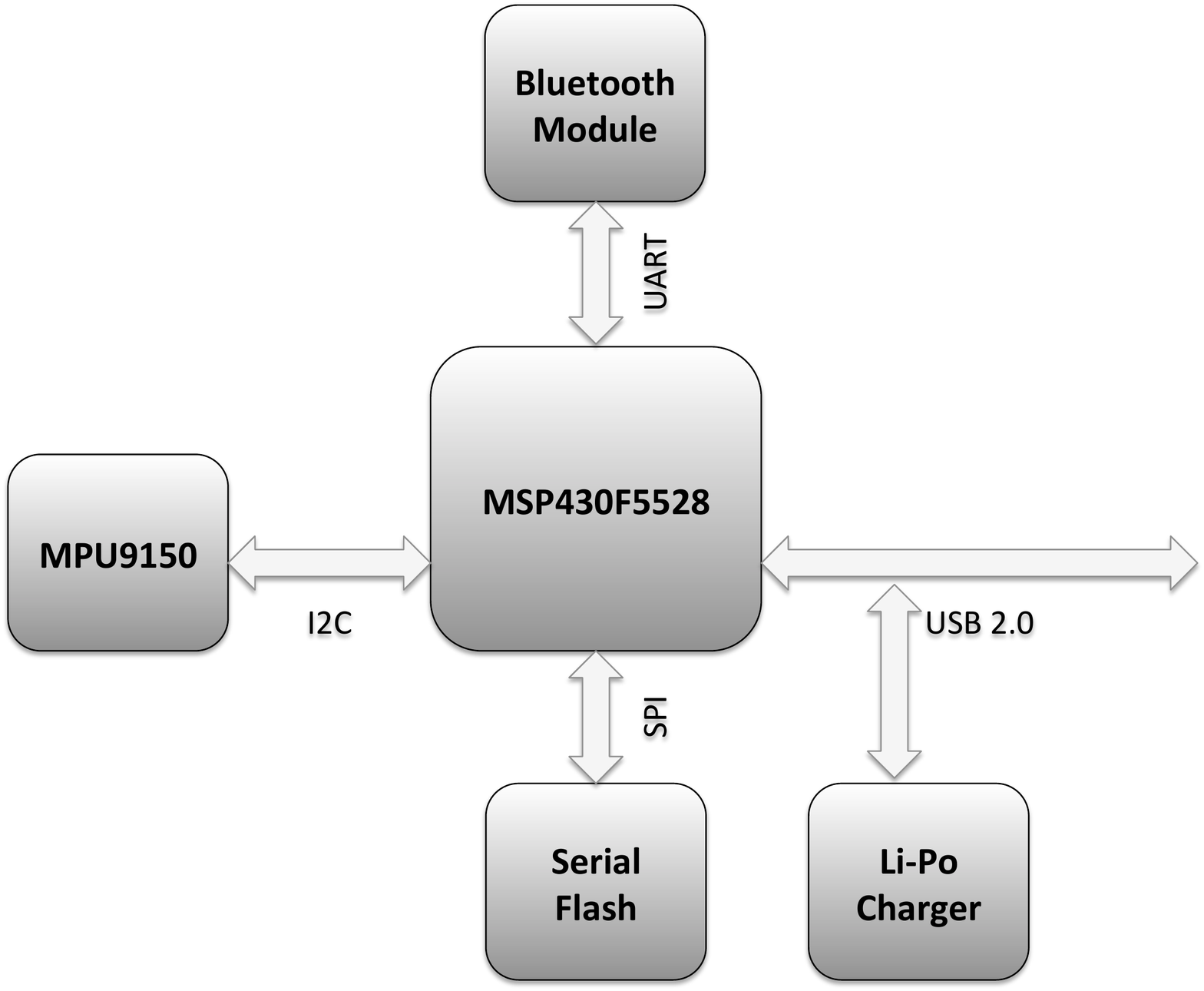}
  \caption{Board components}
  \label{fig:board}
\end{minipage}
\begin{minipage}[b]{0.32\linewidth}
  \includegraphics[scale=0.23]{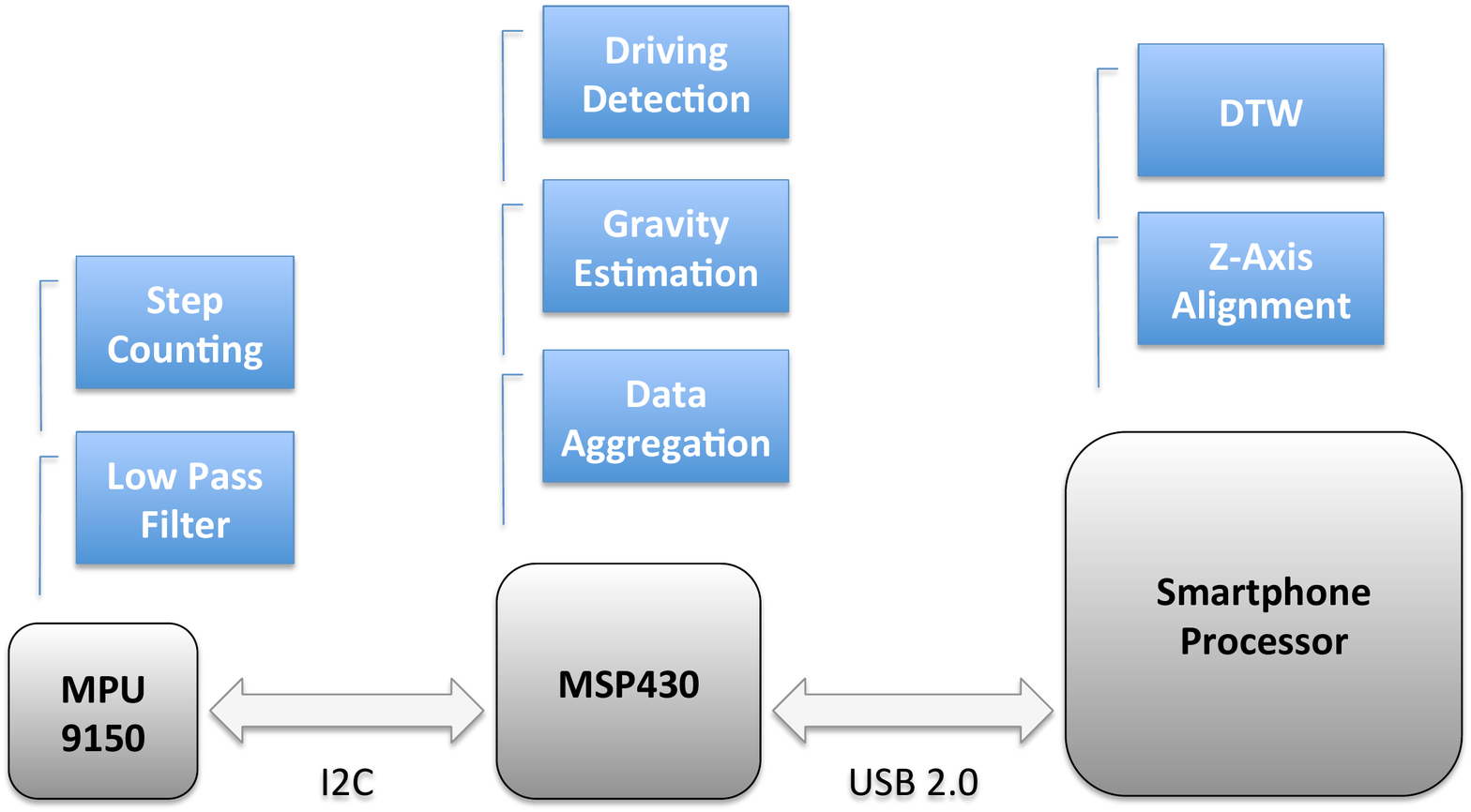}
  \caption{System overview}
  \label{fig:system}
\end{minipage}
\end{figure*}

As we mentioned, the coprocessors on recently released phones are not programmable so to test the performance of our system with a coprocessor we resorted to use different programmable hardware shown in Fig. (\ref{fig:sensorboard}).

Our embedded sensor board is built around a $16$bit MSP430F5528 processor with $8$kB SRAM and $128$kB internal program flash. It provides a rich collection of interfaces including I2C and SPI buses, UARTs and a built in USB $2.0$ interface with its separate $2$kB SRAM. The processor also includes a $16$bit hardware multiplier. A bluetooth $2.0$ module is connected to one of the UARTs of MSP430. The processor is also connected via I2C to a sensor chip MPU-9150 that includes motion sensors. It also has an internal buffer to temporarily store sensor data and generates interrupts to MSP430 to indicate when the data is ready allowing the processor to enter into low power sleep modes between samples. Each of the motion sensors can be individually switched on and off and sampled at rates varying from $10$Hz to $100$Hz. Fig. (\ref{fig:board}) shows a block diagram of our embedded sensor board. This sensor board is connected to a Samsung S2 phone through an OTG USB cable allowing an Android application to communicate with it through the Android USB framework. Fig. (\ref{fig:system}) shows an overview of the entire system with different software and hardware components. Here we describe these components in detail.


{\bf Sensor} We use MPU-9150 as our main sensor. It integrates a high quality accelerometer, a gyroscope and a magnetometer and outputs $16$bit data for each of these sensors at configurable sampling rates. As we mentioned earlier, MEMS sensor data usually contains high frequency noise and requires filtering before it can be used. The chip has an internal digital low pass filter with a configurable cut-off frequency that frees up the processor from digital filter computations. We set up this filter with a cut-off frequency of $42$Hz. The chip also includes a pedometer to detect if a user is walking and step counting. This can be used to detect if a user has finished his or her journey and has started walking. We, however, use a simple activity recognition algorithm running on MSP430 to detect these transitions.

{\bf Low Power Processor} MSP430 runs a simple activity recognition algorithm based on a decision tree to detect when a user starts and stops driving. When it detects that the user is driving, it samples, aggregates and buffers sensor data from MPU-9150, estimates the gravity vector and pushes aggregated data to the phone if realtime route recognition is required. Otherwise the data is stored in the serial flash and can be used later for route mining.  

\begin{table}[t]
\centering
\begin{tabular}{lll}
        & Predicted &      \\
Actual  & Driving   & Null \\ \hline
Driving & 49.7\%      & 0.7\%  \\
Null    & 1.5\%       & 48.1\%
\end{tabular}
\label{tab:confusion}
\caption{Confusion matrix for driving detection}
\end{table}

\begin{figure}[t]
\centering
\includegraphics[scale=0.35]{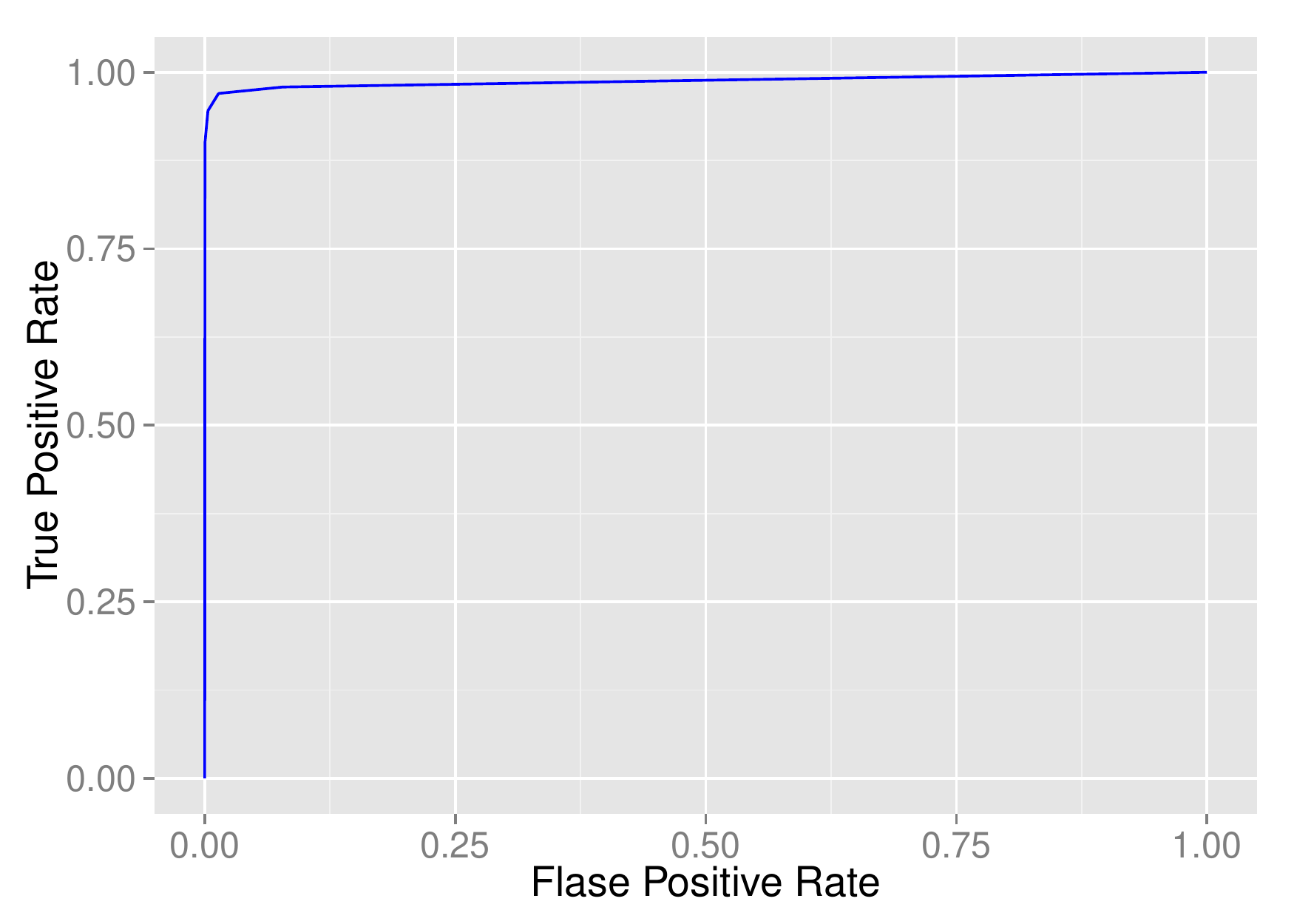}
\caption{ROC curve for driving activity recognition}
\label{fig:roc_curve}
\end{figure}

There is a significant amount of work on activity recognition on smartphones and wearable devices~\cite{activity1, activity2, activity3, transportmodes}. Recent devices like Apple's iPhone 5s already include this capability. Our aim over here is just to demonstrate that, by using one of these techniques, it is possible to perform continuous sensing and to trigger route sensing automatically without user interaction in an energy efficient manner. In order to train a decision tree for driving detection, we collected accelerometer data from the phones of two users who do not use any motorized transport. We instrumented the phones to sample the accelerometer for $8$s after every two minutes and save the data on the phone. This data was collected continuously over $5$ days to capture the entire range of activities that a user might perform apart from driving. This formed the null class for our model training. We used the accelerometer data from the dataset discussed in Section(\ref{sec:route_mining}) that was collected during driving as the primary class. Using these two labelled classes, we trained a simple decision tree with per axis mean and variance of acceleration as features. We split the dataset in a $60\%$ training and $40\%$ testing data making sure that both classes are represented equally in both data partitions. We trained the classifier on the training data and then validated its performance on the testing data. Table (\ref{tab:confusion}) shows the confusion matrix when we run the trained classifier on the testing data never seen by the classifier during training. It shows that both classes are equally represented in the testing data and that the classification error is very small. Fig. (\ref{fig:roc_curve}) shows a ROC curve for this classifier. 

MSP430 samples the accelerometer (the gyroscope is disabled) at $10$Hz, computes features over a $2$s window and runs the trained decision tree driving detection. When driving is detected, it configures the MPU-9150 to sample from both the accelerometer and the gyroscope at $50$Hz. It computes a running average over $1$s on the gyroscope data and after one second transfers the mean $x$, $y$ and $z$ angular speed components to a separate buffer. We use a $1440$ bytes buffer to store up to $4$ minutes of angular speed data. A running mean of the accelerometer data is also calculated over these $4$ minutes to estimate the gravity vector. MSP430 then enables its USB interface and sends buffered angular speed data and gravity vector to the phone. The USB interface is then disabled to save energy.

\begin{figure}[t]
\centering
\includegraphics[scale=0.35]{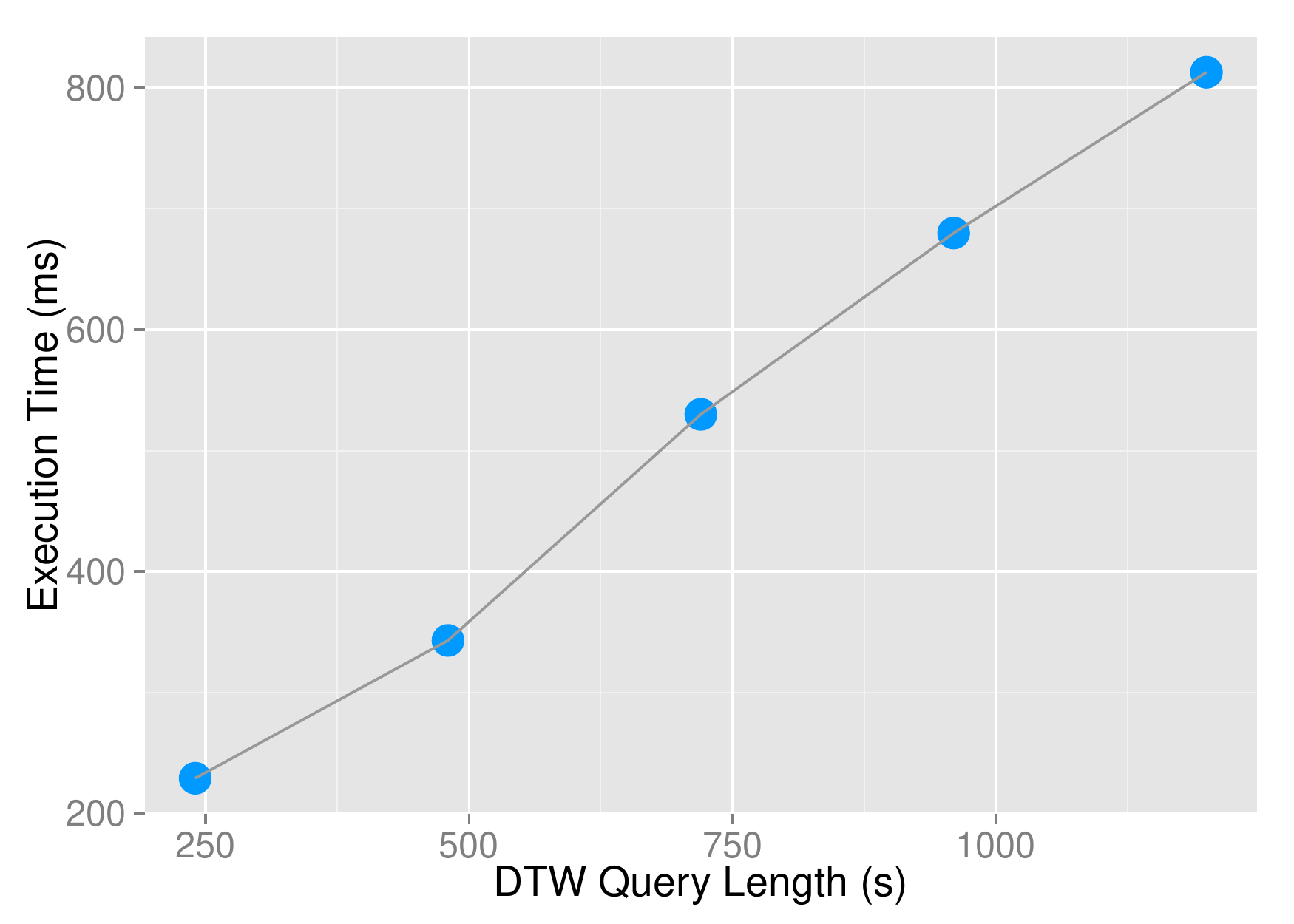}
\caption{DTW execution time on Samsung S2}
\label{fig:dtwexectime}
\end{figure}

{\bf Smartphone Processor} On the smartphone, we developed an application that registers itself with the Android OS with MSP430's USB vendor identifier (VID) and product identifier (PID). When the MSP430 enables its USB interface, a Broadcast Intent triggers Android OS to start our application on the phone even if the phone is in idle standby state. We implemented the Z-Axis alignment discussed in Section ({\ref{sec:zaxis_alignment}}) and Dynamic Time Warping discussed in Section (\ref{sec:dtw}) in this application. It transfers the data from the sensor board to the phone and then either saves it to storage or runs these algorithms to perform realtime route recognition on the phone. When configured to perform realtime route recognition, it executes Z-Aixs alignment and then computes normalized DTW distance between current angular speed trace and cluster centers in its database. As new angular speed data is received from the sensor board after every $4$ minutes, it is combined with the previously received data and the DTW distance is recalculated. This is only done up to $5$ times ($20$ minutes of angular speed data) and if no match is found with any of the saved cluster centers, it is classified as new route not seen before and marked for storage. Fig. (\ref{fig:dtwexectime}) shows the execution time for calculating DTW distance of current angular speed data as it is received from the sensor board with a previous cluster center representing a $24$ minute journey. It shows that execution time increases linearly as new data arrives in from the sensor board and the DTW query length increases. It takes slightly more than $800$ms to compare maximum query length of $20$ minutes with a  cluster center. Z-axis alignment can be executed on each batch of data individually without combining it with the previous batch. On Samsung S2, our quaternion based z-axis alignment executes on average in $4$ms for $4$ minutes worth of angular speed data.

\begin{figure*}[t]
\centering
\begin{minipage}[b]{0.3\linewidth}
  \includegraphics[scale=0.3]{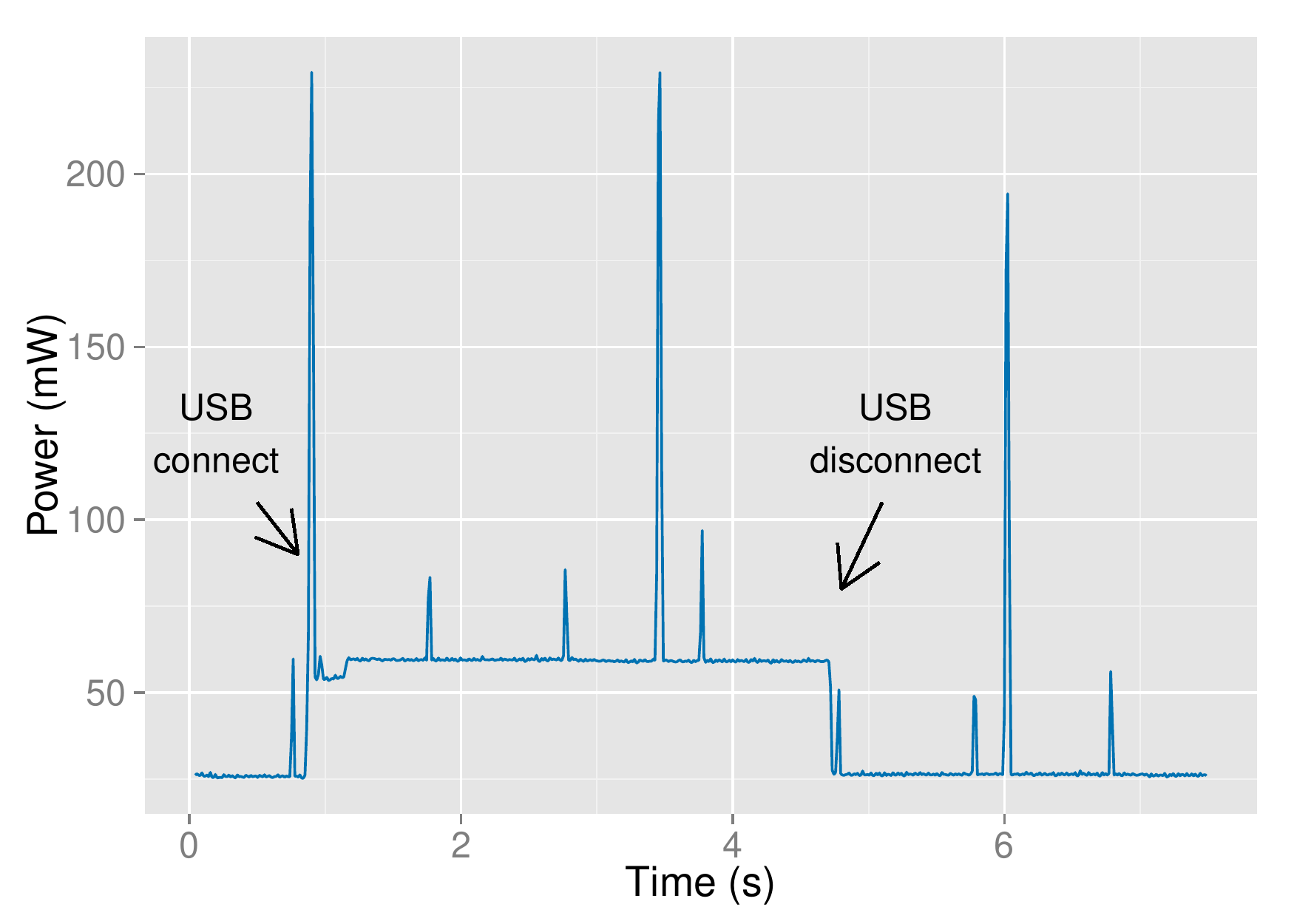}
  \caption{USB transaction}
  \label{fig:boardusb}
\end{minipage}
\begin{minipage}[b]{0.3\linewidth}
  \includegraphics[scale=0.3]{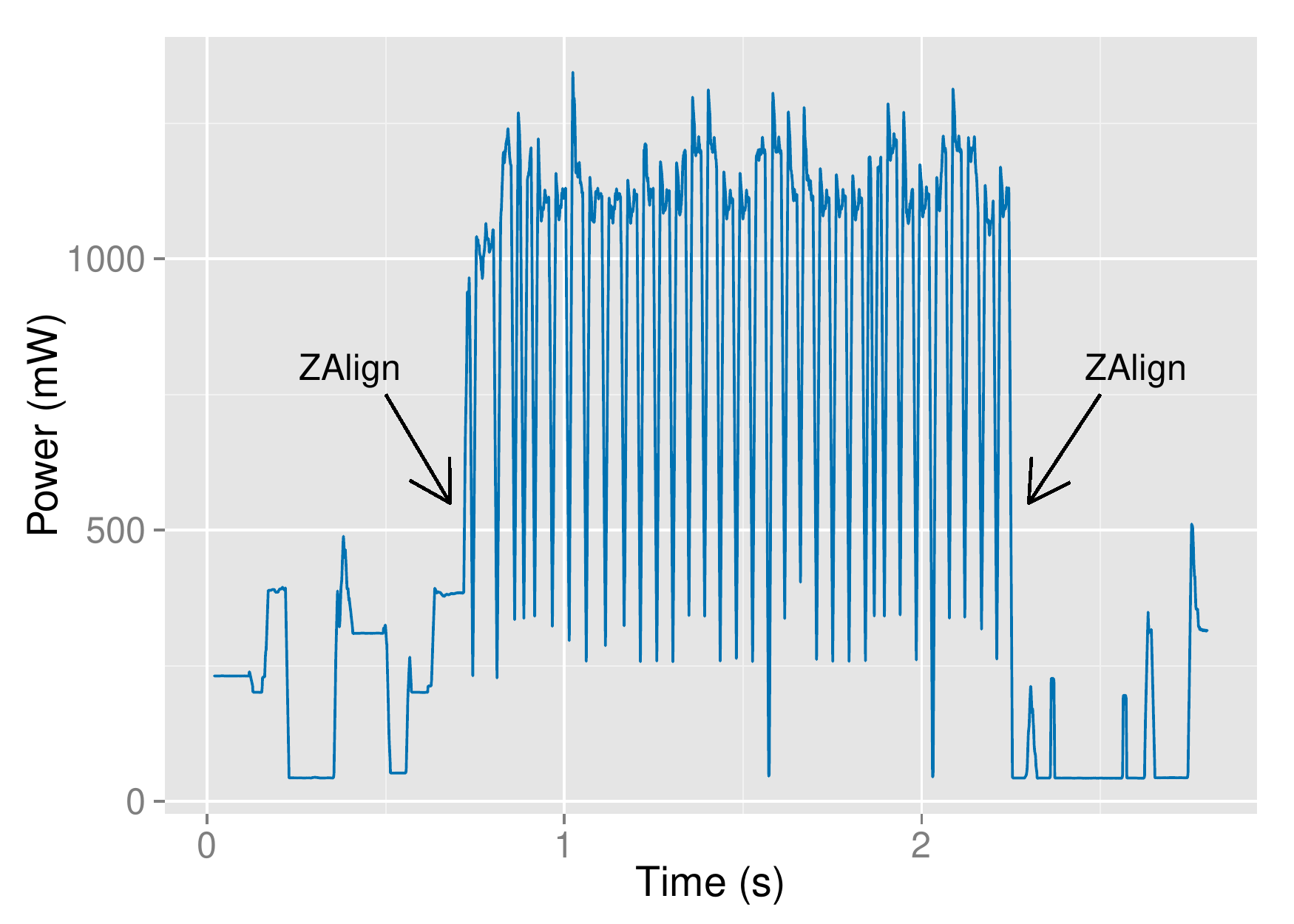}
  \caption{Z-Alignment}
  \label{fig:zpower}
\end{minipage}
\begin{minipage}[b]{0.3\linewidth}
  \includegraphics[scale=0.3]{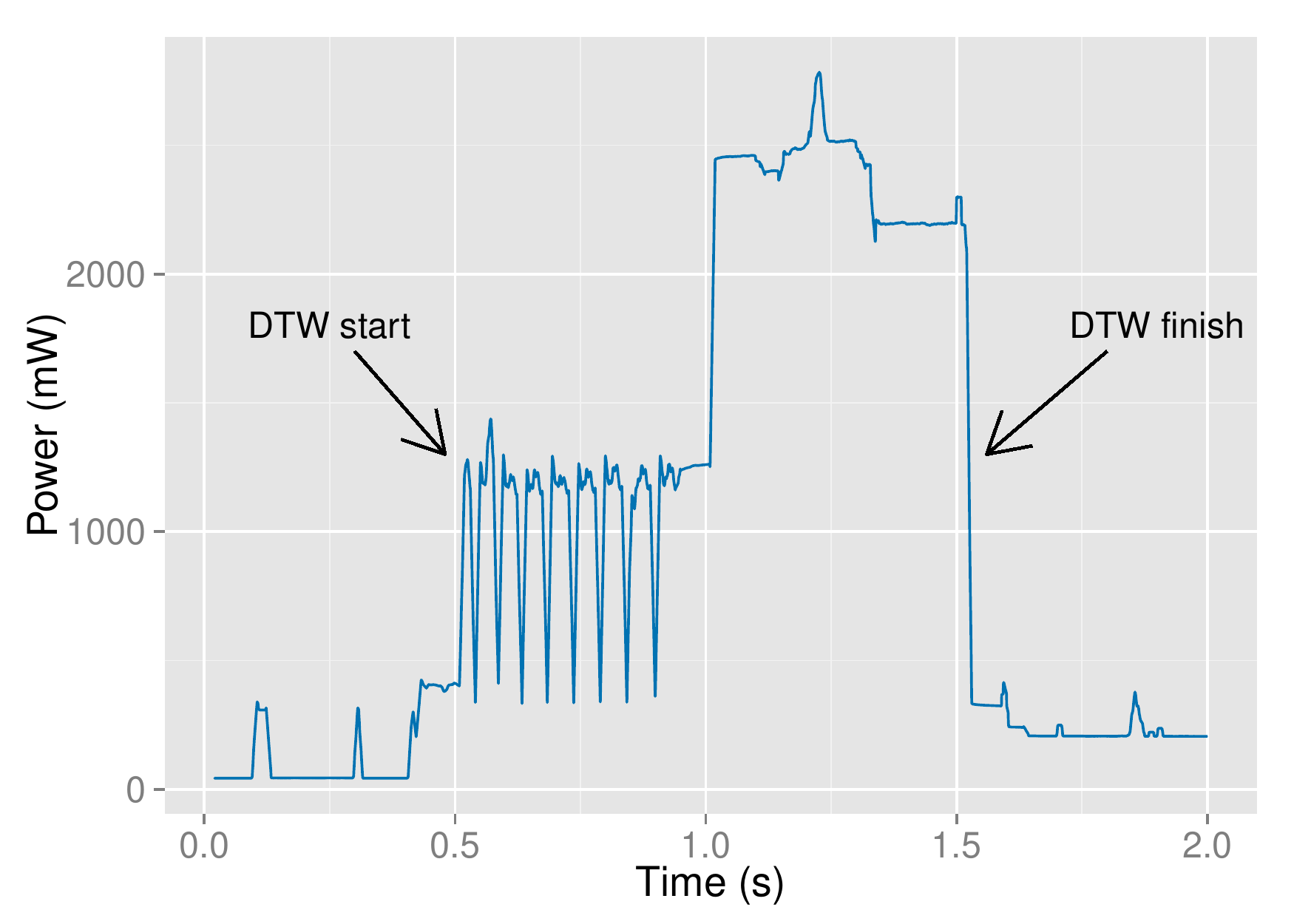}
  \caption{DTW computation}
  \label{fig:dtwpower}
\end{minipage}
\end{figure*}



{\bf Energy Consumption} We now investigate the energy consumption of our route sensing and recognition system. We use the Monsoon power monitor to measure current and power consumption of the system. With the driving detection classifier running on the embedded sensor board, MPU-9150 generates a $10$Hz interrupt triggering the MSP430 to exit the LPM0 low power mode to retrieve the accelerometer data from the chip and to execute feature extraction and classification. In this sensing state, the sensor consumes $5$mW of power. This is less than one third of the energy consumption of Samsung S2 phone in its idle state and two orders of magnitude smaller than sampling motion sensors or GPS on the phone. When the sensor board detects that the user is driving, it enables the gyroscope and configures a sampling rate of $50$Hz for the both the accelerometer and the gyroscope. With data aggregation and buffering, gravity estimation and the activity classifier, the sensor board consumes $25$mW of power. However, now after every $4$ minutes, it enables its USB interface and transfers buffered data to the phone. Fig. (\ref{fig:boardusb}) shows a power trace captured with the power monitor for one of these USB transfers. It shows that power consumption increases to $59$mW when the USB interface is enabled and drops back to $25$mW when the transaction is complete and the interface is disabled to save energy. It lasts for about $4$s mainly because the sensor board has to wait for the phone to load the Android application to receive data. Although, our sensor board has a bluetooth module, we decided not to use for data transfer it due to the its high energy consumption. 



When $4$ minutes of buffered angular speed data and gravity vector is transferred to the phone, z-axis alignment is executed on the phone and takes on average $4$ms on Samsung S2. In order to capture the power consumption of this computation on the power monitor, we repeat it $500$ times in a tight loop on the phone. Fig. (\ref{fig:zpower}) shows the power consumption of Samsung S2 during this loop repeating z-axis alignment computation for $500$ times. It shows that the mean power consumption of z-axis alignment is about $900$mW. Fig. (\ref{fig:dtwpower}) shows the power consumption of Samsung S2 when executing DTW with two $20$ minute long routes i.e. $1200$ angular speed data points (1 per second) for each route. This is the maximum length DTW query that our system ever executes. Any routes not matched within $20$ minutes of driving can safely be assumed to be new routes not seen before and therefore do not require DTW based comparison. Fig. (\ref{fig:dtwpower}) shows that the mean power consumption of Samsung S2 for this task is about $1500$mW. 


Table (\ref{tab:power}) lists the power consumptions of different system components along with their execution times and intervals if any apply to them. The row marked as Sensing refers to the power consumption of the sensor board when it is in route sensing state. This lasts throughout the route and therefore has no execution time or interval associated with it. USB transfers occur every $4$mins ($240$s) and last only for $4$s. Thus the mean power consumption of this task over the entire duration of the route is $0.97$mW. Similarly the mean power consumption of z-axis alignment over the entire duration of route is $0.02$mW. DTW executes only up to a maximum of $5$ times during the entire route. The total execution time however depends on the amount of buffered data received so far and the number of clusters to compare with.  Based on our dataset, we assume a total of eight clusters which gives us a running time of $2$ to $6.6$s as the query length grows from minimum to maximum. The mean power consumption of DTW execution over the entire duration of a $20$min route is thus about $28$mW. For longer routes, the mean power consumption will be lower than this as the cost of running DTW is averaged over the entire duration of the route. The total average power consumption of the complete route sensing and recognition (sum of all components) is about $54$mW.

\begin{table}[t]
\begin{tabular}{lllll}
Function     & \begin{tabular}[c]{@{}c@{}}Power\\ (mW)\end{tabular} & \begin{tabular}[c]{@{}c@{}}Execution\\ Time (s)\end{tabular} & \begin{tabular}[c]{@{}c@{}}Execution\\ Interval (s)\end{tabular} & \begin{tabular}[c]{@{}c@{}}Mean\\ Power\\ (mW)\end{tabular} \\ \hline
Sensing & 25         & -                                                            & -                                                              & 25                                                      \\
USB Transfer & 59         & 4                                                            & 240                                                              & 0.97                                                      \\
Z Alignment  & 910        & 0.004                                                        & 244                                                              & 0.02                                                     \\
DTW       & 1560       & 2 - 6.6                                                          & -                                                             & 28.01                                                      \\ \hline
Total        &            &                                                              &                                                                  &                 54                                         
\end{tabular}
\label{tab:power}
\caption{Mean power consumption of system components}
\end{table}

Fig. (\ref{fig:system}) compares the power consumption of our system with the energy consumption of Samsung S2 in its idle state, sampling motion sensors on the phone and the GPS sensor. It shows that the energy consumption of our system (5mW) in its continuous activity recognition state is two orders of magnitude less than sampling either the motion sensors on the phone or the GPS. When our system is in its route sensing and recognition state, the mean power consumption (54mW) is still an order of magnitude less than sampling either of the sensors on the phone. To put this in context,  the mean power consumption of sending a single text message and replying to an email over WiFi is about $300$mW and $432$mW~\cite{poweranalysis} respectively. For a one hour journey, our approach leads to less than $0.01$\% usage of a typical $1650$mAh phone battery. For phone motion sensors or GPS, we only included the energy consumption of sampling the sensor from the main processor without any further computation or calculations. Adding the cost of any post processing or computation will increase the power consumption of any system based on these sensors even further. 

\begin{figure}[t]
\centering
\includegraphics[scale=0.37]{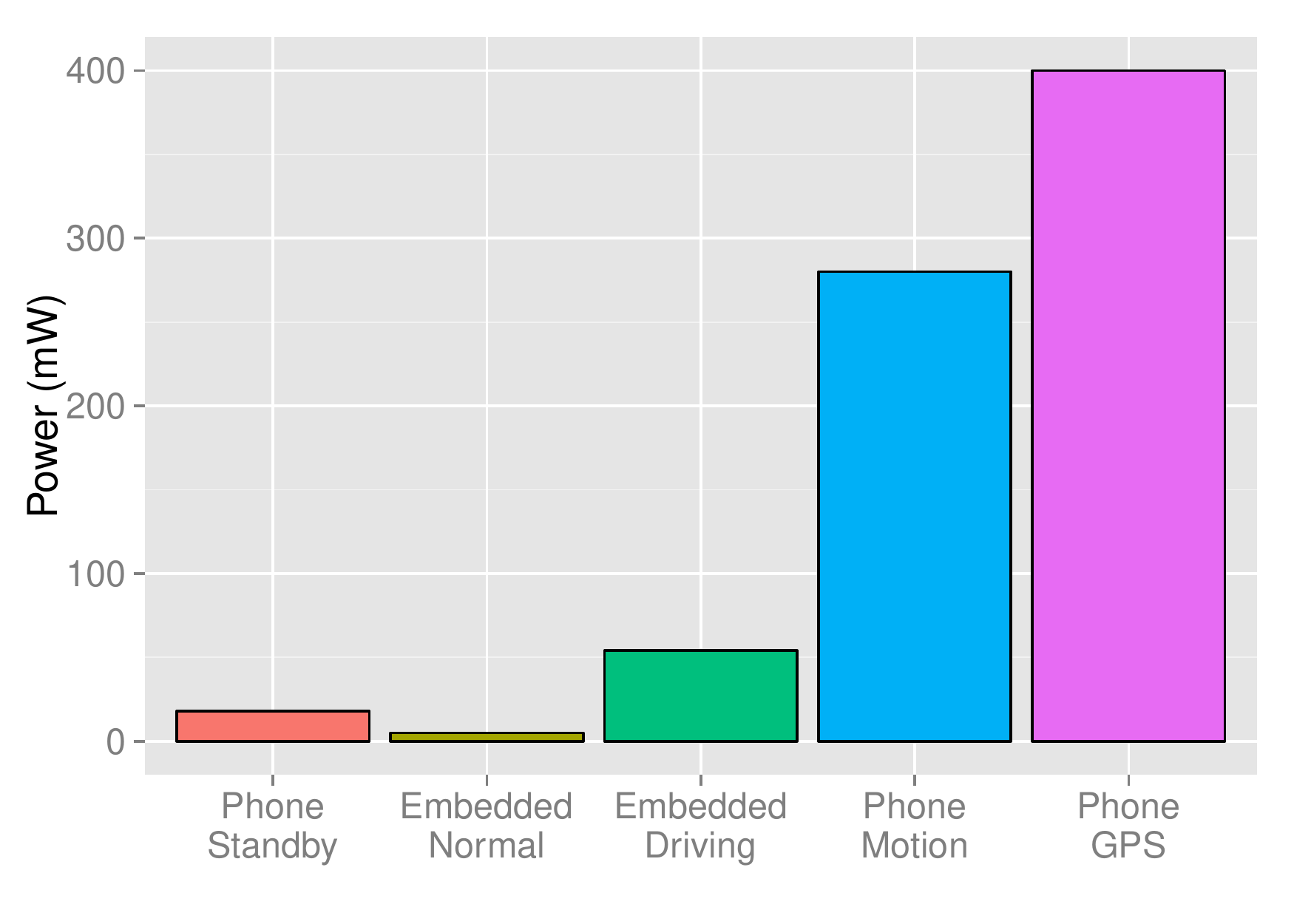}
\caption{Power consumption of our approach compared to phone's motion and GPS sensors}
\label{fig:system}
\end{figure}

\begin{figure}[t]
\centering
\includegraphics[scale=0.32]{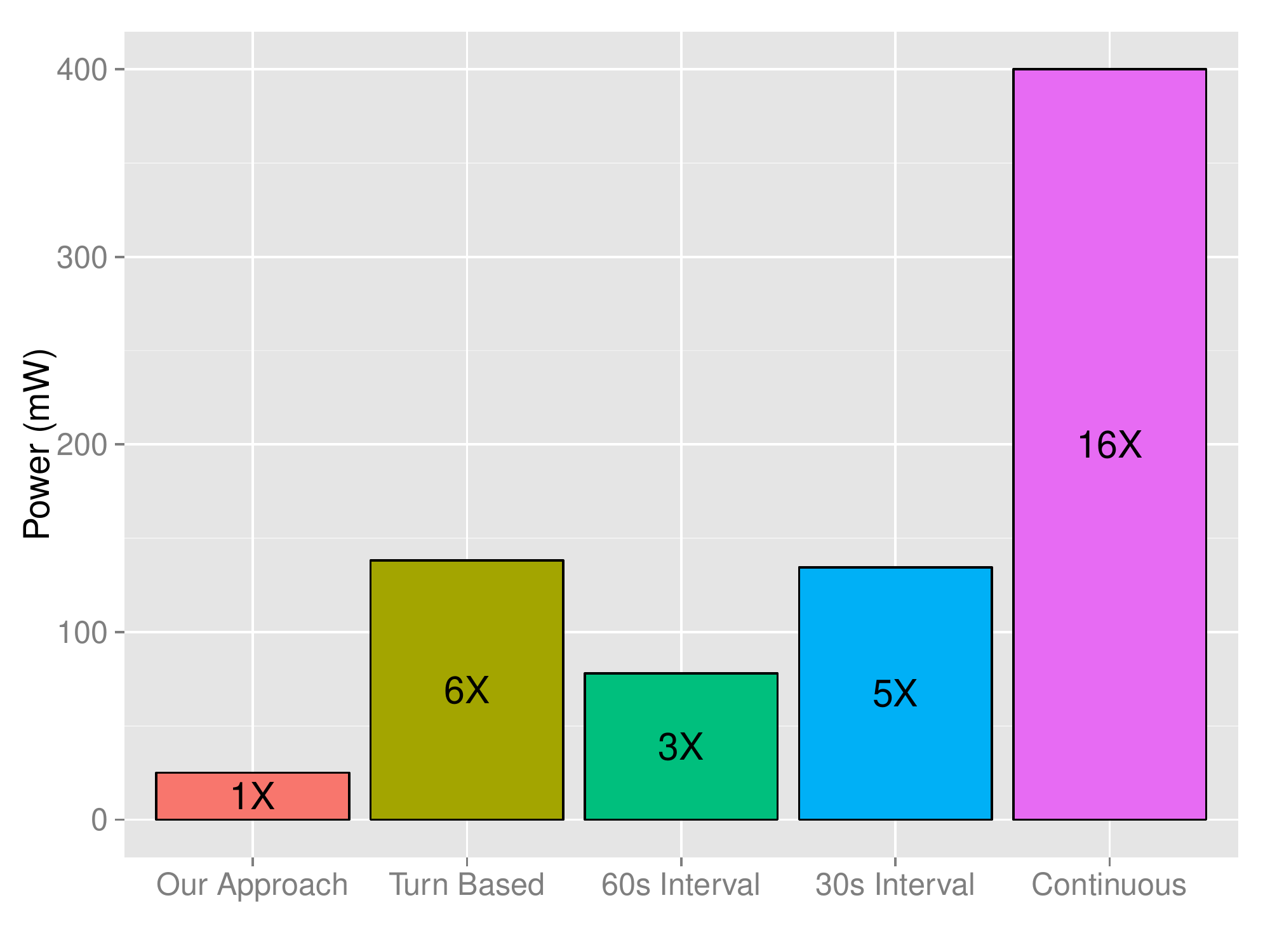}
\caption{Power consumption of our approach as compared to duty cycled GPS}
\label{fig:gpsdutycycle}
\end{figure}

Now we compare the energy consumption of our gyroscope based route sensing (without DTW and route recognition) with that of sampling a duty cycled GPS sensor. Duty cycling a typical GPS receiver is usually not straight forward due to long and unpredictable time required to acquire a fix and loss of hot start state soon after it is switched off~\cite{leap}. Reducing the sampling interval of GPS beyond $30$s in a moving vehicle also makes it difficult to infer the actual route taken by the vehicle with map matching~\cite{mapmatching}. We duty cycle the GPS on Samsung S2 with a $30$s interval while traveling on the path shown in Fig. (\ref{fig:paths}) over a week and measure GPS fix times. The mean acquisition time of GPS in this duty cycled setting is $8$s. We exclude the time required for initial fix and any measurements where the GPS lost the fix after initially acquiring it (which is not uncommon in dense urban areas) thus making the comparison more favorable for the GPS.   Power consumption of GPS in this duty cycled setting measured using Monsoon power monitor is $425$mW. Assuming an initial GPS fix time of $60$s, Fig. (\ref{fig:gpsdutycycle}) compares the energy consumption of our route sensing with both continuous and duty cycled GPS sensor. It shows that even under good conditions (small initial fix time and no loss of signal), energy consumption of the GPS sensor duty cycled with $30$s and $60$s intervals is five and three times of that of gyroscope based sensing. It also shows that a turn based duty cycling approach that detects turns using gyroscope and then triggers the GPS (mean time between turns in our dataset is $27$s) is six times the gyroscope based route sensing. Another approach for reducing GPS energy consumption is to offload GPS signal decoding and location computation to a cloud service~\cite{cogps}, however, these techniques are only suitable for delay tolerant applications and cannot be used (without incurring the data transfer overhead) for real-time route recognition while the user is in transit to trigger various actions e. g. sending notifications to friends/family, switching heating systems etc. 



\section{Discussion}
\label{sec:mapping}


We presented a system that can infer a user's significant journeys by using only the accelerometer and gyroscope of user's phone. Once the system has identified significant routes, it can either infer or gather more information about these journeys. For example, for a typical user, it can infer that the most frequently repeated last route of the day is a journey to home. It can also present a time annotated list of the most frequently taken routes to the user and request a label for these journeys. This approach is quite typical and frequently used by significant place learning systems. For example, Google Now~\cite{googlenow}, a context and place learning application by Google, requests users to mark home and work locations after these have been identified by the system. This label based approach is sufficient for some of the applications that we think will benefit from route learning. For example, a smart home heating application~\cite{tado} can present a list of route labels provided by this system to the user and ask to choose journeys that should start or stop the heating. When any of the selected journeys are detected by the system, it can inform the application with a callback and the application can take necessary actions. Similarly, another application that we envisage is route based alerts or messaging. The user can select a previously labeled route to set up a text message to be sent to family or friends when that journey is detected to inform them that the user is on his way or to send a message when a routine journey is taking unusually long say due to a traffic jam.

There is a second class of applications, however, that require the actual physical route. For example, a traffic alert application that informs the user that there is a disruption on his regular route (an accident or road works etc) requires the actual physical path to provide this type of information to the user. For these applications, our system can annotate the significant routes with location data say from GPS. This approach is still energy efficient as compared to using only location sensing for route learning. It acquires location data that has high energy cost \emph{once only} for those routes that have already been identified as significant as opposed to acquiring location data for all the user journeys only to keep some of it and discarding the rest. This is again something that is used by place learning systems for automatically annotating semantic places with location data.

Another more sophisticated approach that the system can use is to estimate the turning angles by integrating angular speed and measure times between turns for a previously identified significant journey. Given these turning angles, times between turns and the source destination pair (as significant places are already sensed by place learning features on smartphones) of the journey, determine which one of all the possible routes between these two locations was taken by comparing the turns and times between turns from a GIS mapping service like Open Street Map by using a HMM based map matching approach~\cite{autowitness}.

\section{Related Work}

Dynamic Time Warping (DTW) has been used before in a number of settings, even in the sensing domain.
Johnson et. al.~\cite{dtwdrivingstyle} and Eren~\cite{dtwdrivingstyle2} describe a DTW based classifier that can be trained with various driving maneuvers by using the accelerometer and the gyroscope data. They use it to detect the type of the driving maneuver and whether it is being executed in a normal or an aggressive manner to sense the driving style of the user. Unlike this approach we use DTW to match entire routes and not to detect user driving behaviors on a limited set of maneuvers. 

Chandrasekaran et. al.~\cite{dtwvehiclespeed} also use Dynamic Time Warping (DTW) on the radio signal strength obtained from the GSM radio interface of the smartphone traveling in a vehicle. They compare it with a signal profile captured from the same road segment at a known speed to estimate short term vehicle speed variations. Liu et. al.~\cite{gesturedtw} build a library of gesture templates from an accelerometer in a smartphone or a consumer device like Wii Remote and then use DTW for recognizing these gestures. 

Mobile sensing has also been used in several other applications related to vehicles and traveling. For example, a number of studies have used accelerometers in smartphones to monitor road surface conditions~\cite{nericell, pothole}. Others have used phone sensors to monitor driver behavior~\cite{drunkdriving, dtwdrivingstyle, dtwdrivingstyle2}. White et. al.~\cite{wreckwatch} propose to use the accelerometers in phones for automatic traffic accident detection and requesting emergency help for vehicle occupants. Wang et. al.~\cite{vehicledynamics} use the accelerometer and the gyroscope of the phone during driving to detect whether the phone is on the driver side or passenger side of the vehicle. All of these studies indicate that it is possible to use smartphone sensors to sense and monitor the subtle forces experienced by the phones while driving in a vehicle.

Both Android and iOS operating systems have embedded place learning features. Google Now~\cite{googlenow}, the context sensing core of Android OS, learns a user's home and work locations and then occasionally shows travel times between these based on the calculated shortest route. However, there are usually more than one routes available between different locations and which one of these is relevant to the user can only be determined by performing sensing which these applications do not appear to do probably to save energy. Some users have a personal preference or local knowledge and therefore prefer a particular route as opposed to the shortest route. Some users visit other locations, say to pick up a family member, while traveling home.

There is also a large body of research on inertial navigation that uses inertial sensors like accelerometer, gyroscope and magnetometer to estimate and track the position of the moving device by performing sensor fusion in a Kalman or a Particle filter. However, the low cost MEMS inertial sensors used in consumer devices like smartphones exhibit noise characteristics that make them unsuitable for an inertial only navigation. Woodman~\cite{inertialtr} and Tan et. al.~\cite{bumping} show that the position error of an MEMS based inertial navigation system grows to more than $100$m in less than a minute. MEMS based inertial navigation systems, therefore, require periodic external reference information~\cite{bumping, autowitness} to keep this error bounded.

Thiagarajan~\cite{ctrack} use the GSM radio interface of a smartphone along with the embedded accelerometer and magnetic sensor  to estimate phone's trajectory in an energy efficient manner as compared to GPS. This approach, however, requires GSM wardriving data to pre-train the system. Also the software interface to GSM radios is usually  controlled by proprietary drivers that do not expose the complete GSM fingerprint (neighboring cell towers) on most devices~\cite{placelab}. Krumm~\cite{predestination} and Froehlich~\cite{routeprediction} also present destination and route prediction from previously collected trips from GPS receivers in vehicles. Our approach, however, relies on low power sensors in user's smartphone.


\section{Conclusion}
We have proposed an approach which uses solely the accelerometer and gyroscope of the user's phone to detect repeated driving routes. With the advance of modern phones with embedded co-processor technology, our approach is one order of magnitude more efficient than any GPS based solution. Our future work involves generalizing the approach of significant journeys beyond driving to other modes of transport, for example cycles and motor bikes, that do not have a car like constrained movement model and exploring the possibility of implementing all the software components on embedded co-processor.

\section{Acknowledgments}
This research has been funded by the EPSRC  Innovation and Knowledge Centre for Smart Infrastructure and Construction project (EP/K000314).

\bibliographystyle{abbrv}
\bibliography{routesensing}  
\end{document}